\documentclass[useAMS,usenatbib]{mnras}

\usepackage{graphicx}
\usepackage[outdir=./]{epstopdf}
\DeclareGraphicsExtensions{.eps}
\usepackage{pdflscape}
\usepackage{multirow}
\usepackage{numprint}
\usepackage{float}
\usepackage{booktabs}
\usepackage{natbib}
\usepackage{amsmath}
\usepackage{comment}
\usepackage{verbatim}
\usepackage[labelfont=bf]{caption}
\bibpunct{(}{)}{;}{a}{,}{;}
\usepackage{placeins}
\usepackage{tikz}
\def\checkmark{\tikz\fill[scale=0.4](0,.35) -- (.25,0) -- (1,.7) -- (.25,.15) -- cycle;}

\makeatletter
\let\jnl@style=\rmfamily
\def\ref@jnl#1{{\jnl@style#1}}

\newcommand{\high}{H{\sc i}GH}

\newcommand{\citeb}[1]{\citeauthor{#1}, \citeyear{#1}}
\makeatother

\title[Bayesian Estimates of Dust Evolution]{BEDE: Bayesian Estimates of Dust Evolution For Nearby Galaxies}

\author[P. De Vis, S. J. Maddox, H. L. Gomez et al. ]{P. De Vis$^{1}$\thanks{E-mail: pieter.devis1@gmail.com}, S. J. Maddox$^{1}$, H. L. Gomez$^{1}$, A. P. Jones$^{2}$, L. Dunne$^{1}$ \\
$^{1}$ School of Physics \& Astronomy, Cardiff University, Queen's Buildings, The Parade, Cardiff CF24 3AA, United Kingdom \\
$^{2}$  Institut d’Astrophysique Spatiale, CNRS, Université Paris-Sud, Université Paris-Saclay, Bât. 121, 91405, Orsay Cedex, France \\
}
\begin{document}

\date{ 2021 }

\pagerange{\pageref{firstpage}--\pageref{lastpage}} \pubyear{2020}

\maketitle

\label{firstpage}
\begin{abstract}
We build a rigorous statistical framework to provide constraints on the chemical and dust evolution parameters for nearby late-type galaxies with a wide range of gas fractions ($3\%<f_g<94\%$). A Bayesian Monte Carlo Markov Chain framework provides statistical constraints on the parameters used in chemical evolution models. Nearly a million one-zone chemical and dust evolution models were compared to 340 galaxies. Relative probabilities were calculated from the $\chi^2$ between data and models, marginalised over the different time steps, galaxy masses and star formation histories. We applied this method to find `best fitting' model parameters related to metallicity, and subsequently fix these metal parameters to study the dust parameters. For the metal parameters, a degeneracy was found between the choice of initial mass function, supernova metal yield tables and outflow prescription. For the dust parameters, the uncertainties on the best fit values are often large except for the fraction of metals available for grain growth, which is well constrained. We find a number of degeneracies between the dust parameters, limiting our ability to discriminate between chemical models using observations only. For example, we show that the low dust content of low-metallicity galaxies can be resolved by either reducing the supernova dust yields and/or including photo-fragmentation. We also show that supernova dust dominates the dust mass for low metallicity galaxies and grain growth dominates for high metallicity galaxies. The transition occurs around $12+\log({\rm O/H})=7.75$, which is lower than found in most studies in the literature.
\end{abstract}


\section{Introduction}
Galaxies are complex systems consisting of stars, dust, heavy elements and multiple gas phases. These complex systems evolve from relatively simple gas clouds under the influence of ongoing star formation and associated processes. Metals are expelled into the InterStellar Medium (ISM) when these stars end their lives either in the wind of AGB-stars or when they explode as supernovae (SN). Some of the metals that are expelled at the end of a star's life condense as dust grains, which are also mixed into the ISM. These dust grains absorb and scatter about one quarter of the stellar radiation \citep{Bianchi2018} and re-emit most of the absorbed energy at FIR-submm wavelengths. They thus strongly influence the way we observe galaxies. Dust grains also act as a catalyst for the formation of molecules, and metals continue to accrete onto the dust grains in the dense phases of the ISM (see \citeb{Galliano2018} for a review). In the local Universe, dust contains roughly one quarter \citep{DeVis2019} to one half \citep{Maiolino2019} of the heavy elements in the interstellar medium. It is thus important to study both dust and metals when trying to understand the chemical evolution of galaxies. 

Chemical evolution models are a tool that can be used to interpret observed metal abundances and dust masses together with changes in other galaxy properties such as e.g. gas and stellar masses and star formation histories (SFH). This is done by numerically solving first-order differential equations assuming some initial conditions, an initial mass function (IMF), stellar lifetimes, theoretical nucleosynthesis yields for various elements, a star formation (SF) prescription and a prescription for the flow of gas in and out of the galaxy. Chemical evolution models focusing on the metal content of galaxies have been around for a long time, since the pioneering work of \citet{Schmidt1959}, and still form a relevant and active field of study \citep[e.g.][]{Tinsley1980,Carigi2002,Andrews2017,Zhang2018,Romano2019,Kang2021}. Dust can also be included if the proper dust formation and destruction prescriptions are accounted for. Chemical evolution models can be divided into three categories. One-zone models study the integrated properties of galaxies without spatial resolution, and assume instant mixing of dust, gas and metals 
\citep{Dwek1998,Lisenfeld1998,Hirashita2002,Inoue2003,Morgan2003,Valiante2009,Asano2013a,Rowlands2014b, Zhukovska2014,Remy-Ruyer2014,Feldmann2015,Clark2015,DeVis2017b,Popping2017,Zhukovska2018}. 
These models study the changing balance between the dust sources and sinks, and include processes such as the formation of dust in stellar winds and supernovae, dust growth and destruction in the ISM, and radiation field effects. Most recent studies have found grain growth to be the dominant dust source for evolved galaxies \citep[e.g.][]{Dunne2011,Zhukovska2014,Zhukovska2018,DeVis2017b}.

Multi-zone chemical evolution models are similar to one-zone models, but have several regions in which the evolution is tracked separately. These models are often used to study radial gradients of the different properties \citep[e.g.][]{Boissier1999,Spitoni2021}.  Finally hydrodynamical simulations provide the most realistic framework and track the production of metals and dust and how these flow through galaxies \citep{Bekki2013,Bekki2015b,Aoyama2016,Aoyama2019,McKinnon2016,McKinnon2018}. However these hydrodynamical simulations are much more computationally expensive and it is thus not possible to explore the parameter-space in the same way as for one-zone models.

In order to model the build up of metals and dust within the galaxy, it is important to have a realistic prescription of the amount of gas flowing in and out of the galaxy. Pristine gas continuously accretes onto galaxies from the cosmic web as a result of gravity. At the same time gas is blown out of the galaxy driven by the energy released from stars and supernovae \citep{Chevalier1985}, as well as from supermassive black holes/active galactic nuclei \citep{Begelman1991}. Not all of the gas in these outflows will have sufficient energy to leave the gravitational potential well of the galaxy, and will thus eventually fall back onto the galaxy \citep[e.g.][]{Nelson2019}. This recycling of the outflows is often called the galactic fountain. 

There have been numerous studies that have used chemical evolution models including inflows and outflows \citep[e.g.][]{Calura2008,Feldmann2015,DeVis2017b,Zhukovska2018} as well as hydrodynamical simulations that intrinsically include prescriptions for the flow of gas throughout the galaxy \citep[e.g.][]{McKinnon2018,Aoyama2019}. 
More recently, \citet{Irigoyen2020} included a framework to follow the evolution of dust in the atomic, ionised and molecular gas phases.  Typically, in all these studies, there are multiple parameters that can have a similar effect on the observed build up of metals (e.g. changing the IMF or the SN metal yield tables can both affect the metallicity in the models in a similar way). To reveal the degeneracy between such parameters, each combination of parameters has to be explored and evaluated. Additionally, without a statistical framework, even though it is possible to find a model that fits the observations well, it is impossible to know whether this is the only viable option. Without such a framework, one cannot put robust constraints on the input parameters of chemical models. On the other hand, using a formal statistical (e.g. Bayesian) framework, one can determine the most likely values for each parameter and even uncertainties (in case of numerical continuous variables). Bayesian frameworks have been used in previous chemical models to follow the evolution of elements (e.g. \citeb{Rybizki2017}, \citeb{Belfiore2019} and \citeb{Spitoni2021} and references therein). 

\citet{DeLooze2020} used a Bayesian framework to put statistical constraints on a number of dust and chemical evolution parameters. This work provided an improvement on previous works by accounting for a wider range of model parameters yet did not sample the entire parameter-space sufficiently. E.g. it can be seen from their results that the build up of metals with decreasing gas fraction is not modelled well. Additionally, they combine their galaxy observations in six gas-fraction bins before fitting them and focus most of the discussion on two of these galaxy bins only. In this work we use an improved statistical framework which allows us to fit all individual galaxies simultaneously to a combined set of models. Additionally, we show that by sampling a wider parameter-space, a good match to the observations can be found for all late type galaxies. Including realistic models for galaxies spanning a wider range in gas fraction provides stronger constraints as it is often the most unevolved and evolved galaxies (i.e. at the extremes) that provide the most strain on the models. 

In this work, we improve on the model of \cite{DeVis2017b} where we found that low metallicity dwarfs required different dust production parameters to late-type spirals (see also \citealt{Zhukovska2014,Feldmann2015}). We include physically-motivated dust models from laboratory data and more physical outflows to follow dust destruction, formation and recycling in and out of clouds and it tracks both the interstellar and intergalactic medium.  We use a grid of $\sim 40, 000$ models and a MCMC framework to constrain the chemical evolution parameters related to the build up of metals, and to reveal any degeneracy between these parameters. Subsequently we use the same approach to constrain the model parameters related to dust, using the metal-related model constraints from this work as input and comparing models with different dust parameters.  In Section \ref{sec:observations} we present the observational samples that are used to constrain our models. Section \ref{sec:model} details our chemical evolution model and the grid of models used. Our statistical framework is explained in Section \ref{MCMC}. We illlustrate some of the main parameter dependencies of our models in Section \ref{sec:vis} and our results from the statistical framework are presented in Section \ref{MCMCresults}. Finally we list some caveats in Section \ref{sec:caveat} and our conclusions are given in Section \ref{sec:concl}.

\section{Nearby galaxy samples}
\label{sec:observations}
\subsection{DustPedia}
To best constrain our chemical evolution models, it is key to have a sizable sample of galaxies for which we have reliable gas, stellar and dust masses, as well as Star Formation Rates (SFR), metallicities and Nitrogen-to-Oxygen ratios. Few samples have all this data available, and the largest which does is the DustPedia \citep{Davies2017} sample.  
DustPedia is a collaborative focused research project working towards a definitive understanding of dust in the local Universe, by capitalising on the legacy of \textit{Herschel}. The full DustPedia sample consists of 875 nearby (v $< 3000$ km/s), extended ($D25 > 1^\prime$) galaxies that have been observed by \textit{Herschel} and have a near-infrared (NIR) detected stellar component. These galaxies have excellent multi-wavelength aperture-matched photometry available (typically 25 bands; \citeb{Clark2018}). The Spectral Energy distributions (SED) are fitted using CIGALE \citep{Noll2009} and the resulting galaxy properties are presented in \citet{Nersesian2019}. DustPedia uses the physically motivated (based on laboratory data) THEMIS dust model \citep{Jones2013,Kohler2014,Ysard2015,Kohler2015} 
as reference dust model, which was incorporated into CIGALE. 

The gas masses, metallicities and Nitrogen-to-Oxygen ratios for DustPedia are taken from \citet{DeVis2019}. The authors compiled HI fluxes from the literature and combined them with H2 measurements from \citet{Casasola2020}. The total gas masses used include both HI and H$_2$ (either measured or estimated from \citeb{Casasola2020}) as well as elements heavier than Hydrogen\footnote{The correction factor to account for heavier elements is metallicity dependent and ranges from $\xi=1.33$ for zero metallicity to $\xi=1.39$ for solar metallicity.}.  
For the metallicities a literature compilation of emission line fluxes was done and combined with MUSE spectrophotometry. Characteristic metallicities were determined for each galaxy by fitting radial profiles to the available HII regions using a Bayesian framework. Various metallicity calibrations are available, and the \citet[][hereafter PG16]{Pilyugin2016} calibration was chosen as the reference calibration in this work, following \citet{DeVis2019}. We also give results for the IZI calibration in Appendix \ref{IZI} for comparison. Nitrogen-to-Oxygen ratios are also available from PG16. In total, 317 DustPedia sources have all the necessary observations and uncertainties (in the literature compilations, not all sources had uncertainties). Galaxies without uncertainties are discarded as they cannot be used as reliably in our statistical framework. In line with \citet{DeVis2019}, the DustPedia sample has been divided into late-type galaxies (LTG) and early-type galaxies (ETG). Only LTG galaxies are used in the statistical framework in Section \ref{MCMCresults} as our chemical evolution model is not representative for ETGs. Both LTGs and ETGs are shown throughout the plots in this work for completeness. 

\subsection{H{\sc i} and dust selected samples}
Since DustPedia requires a $5\sigma$ detection in the WISE W1 band and a diameter ($D25 > 1'$) as  the selection criteria, it is somewhat biased against dwarf galaxies. In \citet{DeVis2017a,DeVis2017b} we have shown the importance of including unevolved dwarf galaxies when studying dust and gas scaling relations. Given that our aim is to study dust in chemical evolution models, it is crucial that we have observations that span as wide a range of evolutionary states as possible. Therefore we add the HAPLESS \citep{Clark2015} and \high\ \citep{DeVis2017a} samples to increase our sample size at the high gas-fraction end. 
HAPLESS is a blind dust-selected, volume-limited sample of 42 local ($z<0.01$) galaxies detected at 250\,$\mu$m from the H-ATLAS Phase 1 Version-3 internal data release, covering 160\,sq. degrees of the sky \citep{Valiante2016,Bourne2016}. The H{\sc i}-selected \high\ sample is extracted from the same H-ATLAS area and includes 40 unconfused H{\sc i} sources identified in the H{\sc i} Parkes All Sky Survey (HIPASS, \citealt{Barnes1992,Meyer2004}) and the Arecibo Legacy Fast ALFA Survey (ALFALFA, \citealt{Giovanelli2005,Haynes2011}, Haynes et al. {\it priv. comm.}); 24 of these sources overlap with the HAPLESS sample. \citet{DeVis2017a} compiled FUV-submm photometry for each of these samples, and subsequently derived dust masses, stellar masses and star formation rates consistently using {\sc magphys} \citep{daCunha2008}. For consistency with the THEMIS dust masses for DustPedia, the \high\ and HAPLESS dust masses are scaled down by a factor of 1.075 \citep{DeVis2019}. PG16 Metallicities for \high\ and HAPLESS are taken from \citet{DeVis2017b}. Nitrogen-to-Oxygen ratios were calculated for this work following exactly the method used for DustPedia \citep{DeVis2019}. We have removed all the \high\ and HAPLESS sources that are already present in the DustPedia sample and all sources that do not have all the required data available. The remaining 18 \high\ sources and 5 HAPLESS sources are all LTGs and are added to our observational sample.

\section{Chemical evolution model}
\label{sec:model}
\subsection{Build up of metals}
We use a chemical evolution model to build a consistent picture of how the metal, stellar and gas content change as galaxies evolve. We use a one-zone model where only the integrated properties of the galaxies are modelled. We do separate the ISM into clouds and the diffuse ISM. Within their phases, the gas, dust, stars and metals are assumed to be perfectly mixed. This model is directly based on that of \citet{DeVis2017b}, though with considerable changes (especially to the outflow and star formation prescriptions). All models start as pristine clouds of gas, which are converted into stars as a result of the ongoing star formation and a given IMF. The stars are then tracked throughout their lifetimes and when they end their life either as AGB stars or SN, they expel dust and metals into the ISM. Inflows and outflows also alter the ISM content of the galaxies in our model. 

To determine the total stellar, metal, dust and gas content of the model galaxy, it is necessary to integrate over time, as well as over stellar mass. We use a numerical integration for this, with discrete stellar mass and time steps. For the integration over stellar masses, 500 steps are used, logarithmically spaced between the upper and lower limit of the IMF (see Equation \ref{eq:gasmass}). The size of the time steps is set to 30 million years. This value is chosen to correspond to the time for the dissociation of a molecular cloud \citep{McKee1989}, which simplifies the treatment of the cloud dissociation, as will be discussed in Section \ref{subsec:dustparams}. Throughout this work, we show continuous integrals in the equations, corresponding to the theoretical dependencies. In practise these are all implemented using a numerical integration over the discrete steps (i.e. $dt=30\, Myr$).  

The evolution of the gas mass can be described by:

\begin{align}
\frac{d(M_g(t))}{dt}= - \psi(t) + \int_{m_{t}}^{m_U}&\bigl(\left[m-m_{R}(m)\right]
\psi(t-\tau_m)\phi(m)dm  \nonumber \\
     & \mbox{} + I(t) - O(t) + R(t),
\label{eq:gasmass}
\end{align}
where $M_g$ is the gas mass, $\psi(t)$ is the star formation rate, $\phi(m)$ is the stellar IMF (normalised so that $\int_{m_L}^{m_U} m \phi(m) dm =1$) and $m_R$ is the remnant mass of a star of mass $m$ \citep{Ferreras2000}.  $m_U$ is the upper mass limit of the stars (which is set to 120 $M_\odot$), $m_L$ is the lower mass limit of the stars (which is set to 0.8 $M_\odot$) and $m_{t}$ is the lowest mass for which a star could have reached the end of its life by time $t$. The lifetime $\tau_m$ of stars with initial mass $m$ is derived using the model in \citet{Schaller1992}. The first term in Equation~\ref{eq:gasmass} accounts for gas consumed during star formation and the second term for how much gas is returned at time $t$ by stars of all masses combined. The third and fourth term, $I(t)$ and $O(t)$ are simple parameterisations of the rate at which gas is contributed or removed via pristine inflows and outflows respectively. Finally the fifth term $R(t)$ gives the rate at which the outflowing gas is recycled (see Section \ref{inout}). Similarly, the evolution of the mass of metals ($M_Z$) is given by:

\begin{align}
\frac{d(M_{Z}(t))}{dt}= \int_{m_{t}}^{m_U}\bigl( & \left[m-m_{R}(m)\right]
Z(t-\tau_m) \nonumber \\ 
&+mp_Z\bigr) \times \psi(t-\tau_m)\phi(m)dm \nonumber \\
    &+ Z_{I}I(t) - Z(t)(O(t)+\psi(t)) + R_{Z}(t).
\label{eq:Zmass}
\end{align}
Here $Z$ is the metal mass fraction defined as $Z=M_Z/M_g$. The first term accounts for metals expelled by stars and supernovae. This includes metals re-released by stars after they die, and newly synthesised metals ejected in winds and supernovae. $mp_Z$ are the metal yields which are taken separately from SN or AGB yield tables taken from the literature. We will explore a range of different yield tables in this work. The inflows in this work have a pristine metallicity of $Z_{I}=0$ \citep{Coc2012} and the outflows use the current metallicity of the galaxy. The last term in Equation \ref{eq:Zmass} again corresponds to the recycled outflows and will be discussed in the next section. To determine the mass evolution of specific elements (e.g. $d(M_{\rm O}(t))/dt$ or $d(M_{\rm N}(t))/dt$) we use exactly the same formula, where each $Z$ is substituted by the appropriate element and where $mp_{\rm O}$ and $mp_{\rm N}$ then give the SN and AGB yields for these elements from the literature tables. Finally for our models we define the model metallicity and Nitrogen-to-Oxygen ratio as:

\begin{align}
12+\log({\rm O/H})&= 12+\log\left( \frac{M_{\rm O}/16}{1.36 M_g} \right), \\
\log({\rm N/O})&=\log\left( \frac{M_{\rm O}/16}{M_N/14} \right).
\end{align}

The factor 1.36 is to account for elements other than Hydrogen in the gas and 16 and 14 are the atomic weights of Oxygen and Nitrogen respectively. Before calculating the $12+\log({\rm O/H})$ and $\log({\rm N/O})$, the oxygen masses are corrected for the amount of oxygen locked up in dust. Following the THEMIS model, we assume the average oxygen content of dust by mass is 23.8\,per\,cent (Jones et al. \textit{in preparation}).
    
\subsection{Inflows and outflows}
\label{inout}
Galaxies continuously accrete gas from the surrounding intergalactic medium (IGM).
This inflow of pristine gas is particularly strong at early evolutionary stages. We use the prescription of \citet{Zhukovska2014}: 
\begin{align}
I(t)=\frac{M_{\rm inf} e^{-t/\tau_{\rm inf}}}{\tau_{\rm inf} (1-e^{-t_G/\tau_{\rm inf}})},
\label{eq:inflow}
\end{align}
where $\tau_{\rm inf}$ is the infall timescale, $M_{\rm inf}$ is the amount of gas falling into the galaxy and $t_G$ is the total amount of time over which this gas is accreted. The infall timescale is set to $\tau_{\rm inf}=2$ Gyr; we have experimented with changing this value, and found very little difference to the results after a few Gyr, as long as $\tau_{\rm inf} \ll \tau_{G}$. $M_{\rm inf}$ is set to half of the total mass of the galaxy $M_{\rm tot}$ (which is a free parameter). This means the galaxy will start out as a gas cloud with mass of $0.5 M_{\rm tot}$ and the same amount of gas will be accreted by inflows. We have also explored models where $M_{\rm tot}$ was left the same, but was divided differently between the primordial cloud and the inflowing material, without much change to the models after a few Gyr. The exact prescription for the inflows makes little difference as long as the majority happens at early evolutionary stages. We can even start with just a cloud of pristine gas, with little effect on the chemical evolution of a galaxy. Throughout the rest of this work we will refer to the total galaxy mass $M_{\rm tot}$ as the sum of the mass of the pristine cloud and the total infalling material (each set to 50\,per\,cent of $M_{\rm tot}$).

Galactic winds driven by the ongoing star formation have a more significant effect. These outflows drive metals and gas from the ISM and slow down the build up of metallicity. We express the rate of outflowing gas relative to the rate of star formation as the mass loading factor:
\begin{align}
\eta(t)=\frac{O(t)}{\psi(t)}.
\label{eq:massloading}
\end{align}
Due to their shallower gravitational potential wells, lower mass galaxies have higher mass loading factors. Yet at the same time, AGNs in massive galaxies efficiently blow out mass from the galaxy and have high mass loading factors too. We base our outflow prescription on mass loading factors taken from the Illustris TNG50 simulation \citep{Springel2018, Marinacci2018, Pillepich2018a, Naiman2018}. For each halo in the simulation, they measure the outflow rate passing various radii with a range of velocities, as a function of redshift. We were provided with a set of tables giving the median mass loading factor for a radius of 10~kpc, as a function of outflow velocity, stellar mass ($M_*$) and redshift (Nelson \textit{priv. comm.}). Then we use a simple bi-linear interpolation to estimate the outflow for any given  $M_*$ and redshift\footnote{In order to obtain the redshift for our models, the age of the galaxy was linked to redshift using the astropy.cosmology
package, assuming each galaxy was formed shortly after the Big Bang.}, and sum over the velocity components. We also impose a strong limit so that no more than 50 \,per\,cent of the gas mass can be blown out of the galaxy in a single (30 Myr) timestep.


To determine what happens to the outflows after they leave the galaxy, we use the outflow velocity distribution to determine the median velocity of each component in each bin. The mean velocity for each bin is then used together with the average total (baryon + dark matter) halo mass profile for each bin from the Illustris TNG100 simulation (extracted using the illustris\_python package; \citeb{Nelson2019}) to determine the (friction-less) ballistic trajectory for each outflow component using a simple simulation. If the mean outflow velocity is larger than the escape velocity, the outflowing gas is lost into the IGM. However, a large fraction of the outflows fall back onto the galaxy after a time $\tau_{\rm rec}$ determined by its ballistic trajectory. We refer to this returned gas as recycled gas and to the combined processes of outflow and recycling as the `galactic fountain'. Before the gas is recycled, a fraction is lost to the IGM. The fraction of gas lost scales with the time the outflow spends in the IGM (i.e. the time before it is recycled) as:
\begin{align}
R(t) &= \int_{t_0}^{t} \rho(t-t_i) \,O(t_i)\, e^{-\epsilon \,(t-t_i)} \,d(t_i), 
\label{recycle}
\end{align}
where $t_i$ is the time at which the outflows are expelled and $t$ is the time for which we are calculating the infall rate of recycled gas. $\epsilon$ is a scaling factor for how much of the gas that would otherwise fall back onto the galaxy is lost due to interactions with the IGM. In our work, $\epsilon$ is set to 0.2\,$\rm Gyr^{-1}$, corresponding to a loss of 18\,per\,cent of the outflow mass per Gyr spent in the IGM. $\rho(t-t_i)$ gives the fraction of the outflows that are falling back onto the galaxy after a time $t-t_i$, and is determined by the velocity distribution of the outflows, and how long it takes for them to be recycled. 
Since we have discretised our outflows into three components with a single outflow velocity for each (we use the mean velocity of that component as described above), $\rho$ will be equal to 1 when $t-t_i=\tau_{{\rm rec},v_{\rm out}}$ and 0 otherwise. Here $\tau_{{\rm rec},v_{\rm out}}$ is the time it takes the outflow to fall back onto the galaxy given the outflow velocity $v_{\rm out}$ and the galaxy's mass profile. 

Equation \ref{recycle} becomes:
\begin{align}
R(t) &= \sum_{v_{\rm out}} O(t-f_{\rm recy}\,\tau_{{\rm rec},v_{\rm out}}) \,e^{-\epsilon \, f_{\rm recy}\,\tau_{{\rm rec},v_{\rm out}}}. 
\label{recycle2}
\end{align}
 Here we have introduced the scaling factor $f_{\rm recy}$ as an additional free parameter in our models. $f_{\rm recy}$ scales the recycling time $\tau_{\rm rec}$ up or down to account for the uncertainty in our calculation of $\tau_{\rm rec}$. The sum in Equation \ref{recycle2} is over the different components $v_{\rm out}$ for which $\rho=1$. Usually this means the three outflow components are summed, but occasionally multiple components from different redshift or mass bins contribute to the current recycling rate. We note that if the outflow velocity is larger than the escape velocity of the halo, $\tau_{{\rm rec},v_{\rm out}} = \infty$ and the contribution to $R(t)$ will be zero. This gas is automatically lost to the IGM. 

The metal and dust content of the outflows is determined by the galaxy's metal-to-gas ($Z$) and dust-to-gas ($\delta$) ratios at the time the gas is blown out ($t-f_{\rm recy}\tau_{{\rm rec},v_{\rm out}}$). For the outflows with $v_{\rm out} > 150\ \rm{km/s}$, it is assumed that all dust is destroyed by shocks \citep{Jones1996} and transformed to gas-phase metals. By the time the outflows are recycled, the galaxy will have typically evolved to higher $Z$ and dust-to-gas ratios, which are then diluted by the recycled gas. The galactic fountain thus slows down the build up of metals and dust. 

\subsection{Star formation rates}
\label{secSFR}
In \citet{DeVis2017b}, we used a few template SFHs (one exponentially declining, one delayed and one bursty SFH) for our models. The delayed SFH was found to be the best prescription for most sources. However for that study, the mass loading factors of the outflows were smaller and not mass-dependent. When the mass-loading factors are mass dependent, as is the case in our work, it is not possible anymore to define one SFH for all models. Galaxies with higher mass-loading factors will run out of gas at different times, and it is not possible to account for this using a single SFH. Instead a better approach is to define our SFR based on a given star formation efficiency ($\rm{SFE}=\psi/M_g$). When using a fixed SFE, if a galaxy runs out of gas (due to the combination of ongoing SF and outflows), the SFR will automatically decrease, following the reduction in available gas mass. As a result the outflow rate will also decrease (mass loading factor stays the same). The resulting SFH will have the shape of an exponentially declining SFH. 

Although a constant SFE is an improvement compared to using one single SFH, it is still not ideal as not all galaxies have the same SFE. Indeed, SFE is likely correlated to the stellar mass of a galaxy as the higher mass surface density produces a higher hydrostatic pressure in the ISM \citep{Elmegreen1989,Wong2002}, it may also be redshift dependent since a higher turbulence at a fixed stellar mass would result in lower efficiencies \citep{Hayward2017}. Finally, it also well known that very low gas fraction sources are usually quenched. 

\begin{figure}
  \center
\includegraphics[width=\columnwidth]{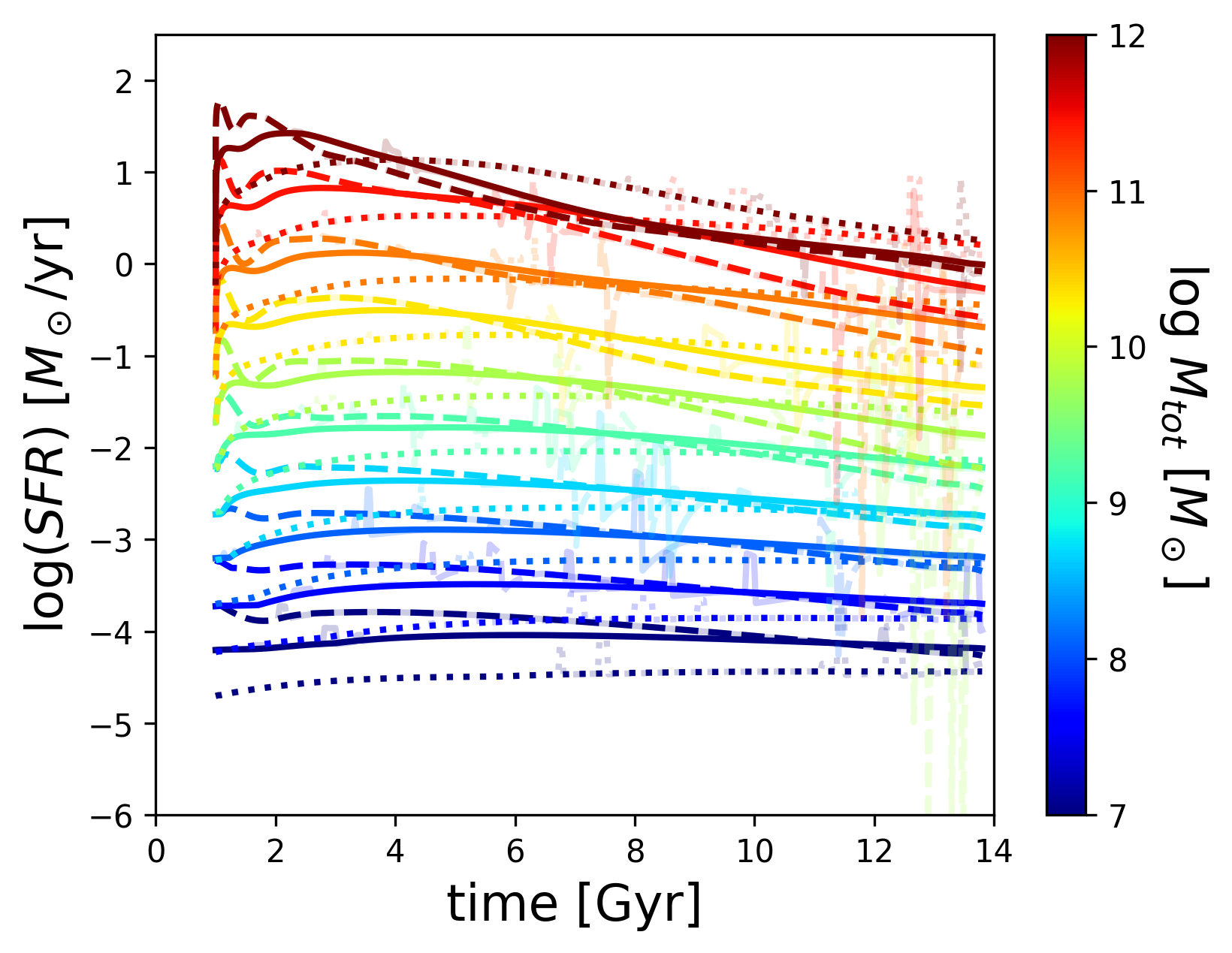}
  \caption{Star formation histories resulting from our empirical SFE prescription for a subset of chemical evolution models (one family of models - see Section \ref{MCMC}). The colour of the model indicates its total galaxy mass, and the linetype the reference SFE ($\rm{SFE}_0=10^{-8.5}\ \rm{yr}^{-1}$: solid, $\rm{SFE}_0=10^{-9}\ \rm{yr}^{-1}$ : dashed and $\rm{SFE}_0=10^{-9.5}\ \rm{yr}^{-1}$: dotted). SFHs including bursts have been included using lighter-shaded lines. The SFHs have the shape of a delayed SFH, which peaks at different times for different mass galaxies. }
  \label{SFH}
\end{figure}

The star formation in this work is made up of (i) a continuous component with a slowly varying star formation efficiency (${\rm SFE}=\psi/M_g$) and (ii) bursts superimposed on the continuous component. Accounting for the dependencies described above, we create a SFE prescription that varies with mass, redshift and gas fraction, and empirically produces SFRs that match the observations (multiple power-law exponents were trialled). The prescription is given by:

\begin{equation}
\begin{aligned}
{\rm SFE} ={} & {\rm SFE}_{0}  \left(\dfrac{M_*}{10^9}\right)^{0.25}  \left(1 + {\rm exp}^{M_*/10 M_g}\right)^{-3}  (1+z)^ {-1},
\end{aligned}
\label{eq:SFE}
\end{equation}
where $M_*$ is the stellar mass, $z$ is the redshift\footnote{The same redshifts were used as for the outflows (see previous footnote).} and $\rm{SFE}_{0}$ is the reference SFE which is a free parameter. Three options of $\rm{SFE}_0$ are explored in this work: $\rm{SFE}_0=10^{-8.5}\ \rm{yr}^{-1}$ (fast), $\rm{SFE}_0=10^{-9}\ \rm{yr}^{-1}$ (average), and $\rm{SFE}_0=10^{-9.5}\ \rm{yr}^{-1}$ (slow). We note that this prescription is not physically motivated, but does provide a sensible framework, where low gas fraction galaxies are quenched, yet immature low mass galaxies also have rather low SFE. The resulting SFH that follow from Equation \ref{eq:SFE} also match the delayed SFHs observed in various works (Figure \ref{SFH}), and are naturally consistent with any ongoing changes to the gas mass.

For the bursty SFH, we follow \citet{Bruzual2003} and generate random bursts so that there is a 50\,per\,cent probability that a galaxy had a burst within any 2 Gyr period of its evolution.  The duration of the burst is randomly drawn from a uniform distribution between 30 and 300 Myr. The SFR during the burst is set so that the stellar mass formed during the burst is a given fraction of the stellar mass at the time. The amount of stars formed for each individual burst is drawn from a loguniform distribution between 0.004 and 0.4 times the stellar mass of the galaxy at the start of the burst. Some examples of the SFH used in this work are shown in Figure \ref{SFH}. After a burst, there is often a period of reduced SFR compared to the continuous SFH. This results from gas being used up (both by star formation and outflows) and since the SFE only varies slowly in our model, the drop in gas mass after the burst results in a drop in SFR. The reduced SFR (and associated outflows), combined with recycling of the outflows expelled during the burst, means that the SFR in a bursty SFH converges back to where it would be for the continuous SFH.

\subsection{Dust parameters}
\label{subsec:dustparams}
Interstellar dust forms in a range of environments, such as the winds of evolved low-to-intermediate mass stars (LIMS, \citeb{Sargent2010}), core-collapse supernovae ejecta (SNe) \citep{Dunne2003,Gomez2012B,DeLooze2017} and grain growth and accretion in the ISM \citep{Mattsson2012}. Some of the produced dust and metals will stay in the molecular clouds they were formed in, and some will dissipate into the diffuse ISM. As the molecular clouds collapse to form the next generation of stars, the newly formed dust will be consumed together with the gas as fuel for the stars. Supernovae shocks also destroy dust as high-energy ions `sputter’ atoms from the surface of dust grains, and collisions between dust grains also break them up \citep{Jones1996,Jones2011}. 
Additional processes such as thermal sputtering, and ionising destruction by cosmic rays, high-energy photons, and free electrons further reduce the dust mass (see \citeb{Jones2004} for a review).

To model the build up and decline of dust, we include dust formation by stars (both LIMS and SN), dust grain growth in the diffuse and dense environments of the galaxy as well as dust destruction by SN shocks and photo-fragmentation of large grains. In this work, we essentially only track the evolution of large grains, as these make up the vast majority of the total dust mass. Small grains are easily destroyed \citep{Bocchio2014} and thus will not be able to build up to a significant amount of dust mass, even though they are essential to explain the MIR range of the SED. 

The different dust processing mechanisms typically affect either only the diffuse ISM or only the cloud ISM. We therefore consider these two phases separately. Note that any reported dust-to-gas ratios will be the total dust-to-gas ratio i.e. total dust mass divided by total gas mass, unless clearly indicated otherwise. $f_c$ gives the mass fraction of the ISM that is in cold dense molecular clouds. This fraction is kept constant throughout the evolution, but as clouds are constantly dissociated and reformed on timescales of $\sim$30 million years \citep[e.g.][]{McKee1989}, we reinitialise the ISM phases after every 30 million year timestep (the total gas and dust mass is re-divided between clouds and diffuse ISM according to $f_c$). We do this by first mixing the cloud ISM into the diffuse ISM. During this cloud dissociation, the newly accreted dust mantled due to cloud grain growth will be exposed to a much harsher environment. As a result the newly formed dust mass is reduced by 90\,per\,cent to account for the evaporation of ices in the dust mantles and to account for the processing of a-C:H mantle material into refractory a-C dust (Jones et al. \textit{in preparation}). After the dissociation of these clouds, new clouds are formed using the updated dust-to-gas ratios from the diffuse ISM.
The evolution of the total large grain dust mass is given by:

\begin{align}
\frac{d(M_{d}(t))}{dt}=& \int_{m_{t}}^{m_U}\bigl(\left[m-m_{R}(m)\right]
Z(t-\tau_m) + mp_Z\bigr) \nonumber \\ 
&\qquad \ \ \times y_d(m) \times \psi(t-\tau_m)\phi(m)dm \nonumber \\
    &+ \delta_{I}I(t) - \delta(t)(O(t)+\psi(t)) + R_{M_d}(t) \nonumber\\
    &+ f_c\ f_{\rm dis}\ M_d \ \tau_{\rm gg,cloud}^{-1}+ (1-f_c)\  M_d\ \tau_{\rm gg,dif}^{-1} \nonumber\\
    &- (1-f_c) \ M_d\  \tau_{\rm destr}^{-1} \nonumber\\
    &- (1-f_c)\, (1-f_{\rm Si}) \ M_d\ \tau_{\rm frag}^{-1}.
\label{eq:Mdmass}
\end{align}

This expression is similar to Equation \ref{eq:Zmass}, with the dust-to-gas ratio ($\delta=M_d/M_g$) in place of $Z$. Noticeably, the astration, inflow, outflow and recycling terms are essentially the same, where $\delta_{I}$ is the dust-to-gas ratio of the inflowing material (set to zero) and $R_{M_d}(t)$ is the dust mass recycled by outflows. At the end of Section \ref{inout}, we saw that only the low-velocity outflow component ($v_{\rm out} < 150\ \rm{km/s}$) contained any dust. Any time dust is blown out of the galaxy in this low-velocity component, we use the ballistic trajectory to determine if and when this dust will re-enter the galaxy as $R_{M_d}(t)$.
 
The first term in Equation \ref{eq:Mdmass} gives the dust that is expelled by stars, which is determined from the amount of metals that is expelled, multiplied by a mass-dependent condensation efficiency $y_d(m)$. For LIMS, this efficiency is fixed at a value of 0.15 and will not be varied throughout this work\footnote{After experimentation we found that changing the LIMS condensation efficiency has barely any affect on the final models. We thus do not include it as one of our free parameters in this work.}. For SN, we use $y_d(m)$ based on the results of \citet{Todini2001}, and add a free parameter to scale the amount of dust produced up or down.

The dust grain growth, destruction by SN shocks and fragmentation rates apply to only one of the two ISM phases (as indicated by $f_c$ or $1-f_c$), and the various $\tau$ give the relevant formation and destruction timescales for which the prescriptions are given below. As can be seen from the $(1-f_c)$ terms in Equation \ref{eq:Mdmass}, our dust destruction by SN shocks and fragmentation only remove dust in the diffuse ISM, as the SN shock velocities get too low to destroy dust in molecular clouds, and self-shielding in clouds makes the photo-fragmentation inefficient. $f_{\rm dis}$ is a factor to account for how much cloud grain growth dust survives the dissociation of the clouds.
$f_{\rm Si}$ is the fraction of the dust that is made up of silicate cores, which are too robust to be affected by photo-fragmentation. 

For the grain growth and destruction terms, we base our prescriptions on the THEMIS dust model \citep{Jones2013,Kohler2015,Ysard2015,Jones2017}, which was also used in the determination of our observed dust masses \citep{Nersesian2019}. The THEMIS model consists of a mix of different grains with carbon and silicate core-mantle structures, for which the properties have been determined from laboratory-measured properties of physically reasonable interstellar dust analogue materials. THEMIS includes carbonaceous and silicate dust and grains of a range of different sizes. In this work we focus on the large grains, which make up the bulk of the mass. The ability of dust grains to evolve in response to the local physical  conditions is one of the key concepts in THEMIS and different environments process grains in different ways. The cores of large grains typically consist of silicates or amorphous carbon (a-C). These cores usually accrete or form photo-processed mantles of amorphous carbon (a-C). In the transition to denser cloud environments secondary a-C:H mantles form (a-C:H stands for hydrogenated amorphous carbon, i.e. the mantles were formed with more hydrogen in their molecular structure). The mantle formation increases both the mass of the individual grains as well as the total dust mass. In these dense clouds, the grains also begin to coagulate into aggregates (the mean grain size increases, but not the total dust mass).  Then, in the the densest cloud environments, ice mantles will form on the aggregate grains (further increasing in the dust mass). 
We include grain growth terms for both the diffuse ($n_H \sim 10^{2}\,\rm cm^{-3}$) and dense ISM ($n_H >  10^{4}\,\rm cm^{-3}$).  Although we expect that the latter will dominate over the former due to the dependence on the accretion timescales with $n_H$, we include both phases here in order to not a priori make any assumptions.

The following prescriptions for the grain growth and destruction timescales were taken from Jones et al. (\textit{in preparation}). We refer to this work for further details on how these formulae were derived. Note that we express our prescriptions in terms of the associated timescales. Each of these enter Equation \ref{eq:Mdmass} as $d(M_{d}(t))/dt \propto M_d \ \tau^{-1}$.
The grain growth timescales are given by:

\begin{align}
\tau_{\rm gg,dif}^{-1} &= k_{\rm gg,dif}\ Z_0 \ \left(1-\dfrac{M_d}{M_Z \times f_{\rm dif}}\right), 
\nonumber \\
\tau_{\rm gg,cloud}^{-1} &= k_{\rm gg,cloud}\ Z_0 \ \dfrac{{\rm SFR}}{M_g} \ \left(1-\dfrac{M_d}{M_Z \times f_{\rm cloud}}\right),
\label{ggcloudeqn}
\end{align}
where $Z_0=Z/Z_{\rm MW}$, i.e. the current metallicity relative to the Milky Way metallicity ($Z_{\rm MW}=0.0134$), $k_{\rm gg,dif}$ is the diffuse grain growth scaling factor in $\rm Gyr^{-1}$ and $k_{\rm gg,cloud}$ is the dimensionless cloud grain growth scaling factor. The $(1-M_d/(M_Z \times f))$ factor accounts for the the depletion of the relevant elements.  $f_{\rm dif}$ and $f_{\rm cloud}$ give the fraction of metals that are available for accretion in the diffuse ISM and molecular clouds. In the dense environments of clouds, the formation of ices and strong accretion of oxygen and carbon becomes possible and $f_{\rm cloud}$ is 2.45 times higher than $f_{\rm dif}$. The SFR is included in the prescription for $\tau_{\rm gg,cloud}$ as stars form in dense ISM regions and this is where the bulk of the available matter that can form dust will accrete into mantles. 
 When the molecular clouds are dissociated, the ice mantles evaporate and only $\approx 10$\,per\,cent of the cloud dust mass accreted in dense clouds survives as a refractory material in the transition into the diffuse ISM \citep{Jones2019}. We set $f_{\rm dis}=0.1$ (Equation \ref{eq:Mdmass}) to account for this.

Our prescription for the dust destruction by SN shocks is very similar to that from \citet{DeVis2017b}. However, to be consistent with the THEMIS framework, it is now expressed in terms of the mass of dust destroyed per SN, $M_{\rm destr}$. The destruction timescale is then given by:
\begin{align}
\tau_{\rm destr}^{-1}  = 135 \ M_g^{-1} \ R_{\rm SN}\ M_{\rm destr},
\label{destreqn}
\end{align}
where $R_{\rm SN}$ is the SN rate. We convert the dust destruction rates for SN shocks from \citet{Bocchio2014} and find a $M_{\rm destr}$ of $\sim30 M_\odot\ \rm SN^{-1}$ and $\sim10\ M_\odot\ \rm SN^{-1}$ for carbonaceous and silicate dust respectively. This produces similar results to the classic destruction model of \citet{McKee1989}, where the dust mass shocked by a SN is proportional to the SN energy, the dust-to-gas ratio and the velocity of the SN shock. Assuming the THEMIS gas-to-dust ratio of 135 and shock speeds of 100 (200)$\rm \, km~ s^{-1}$, \citet{McKee1989} implies 27~(9) $M_{\odot} \rm \, SN^{-1}$ of dust is destroyed\footnote{assuming the relative amount of carbon and silicon is 1:2.}. In this work, we treat $M_{\rm destr}$ as a free parameter, with range of values guided by \citet{Bocchio2014} and \citet{McKee1989}.

Finally we also include the photo-fragmentation of large a-C:H/a-C grains. Silicate grains are too robust to be affected by photo-fragmentation, though the Carbon mantles around them are. Photo-fragmentation is included as a destruction term as this mass is removed from the large grains. The a-C nano-particle dust grains that are formed in this process will be rapidly destroyed and will never amount to a significant fraction of the dust mass (though they often account for a significant fraction of the dust luminosity). The photo-fragmentation timescale is given by:

\begin{align}
\tau_{\rm frag}^{-1} &= k_{\rm frag}' \ G_0, \nonumber \\
&= k_{\rm frag} \ \rm SSFR.
\label{frageqn}
\end{align}

Here we have assumed the diffuse UV radiation field $G_0$ is proportional to the specific star formation rate SSFR. This assumption can be made since a higher SFR per unit stellar mass would result in a larger contribution of UV photons from (hot) young stars, which in turn determines $G_0$. The corresponding scaling factor between these has been folded in when going from $k_{\rm frag}'$ to $k_{\rm frag}$.  We set $f_c=0.5$, $f_{\rm dis}=0.1$ and $f_{\rm Si}=0.1$, consistent with average ratios in the THEMIS model. 

\subsection{Grid of models}
\label{sec:modelgrid}
In order to better understand the parameter space, we build grids of models and compare to the observed properties of our galaxy samples. We vary the key parameters that determine the chemical and dust evolution of our models. The values and symbols we have used are listed in Table \ref{gridtable}. We have made models with:

\begin{table*}
\caption{Free parameters in the chemical evolution models together with the grid values used to sample the parameter-space.}
\centering
\footnotesize
\begin{tabular}{l|ll} \hline\hline
Parameter & values & Notes\\ 
 \hline
$\log\ M_{\rm tot}/M_\odot$ & 7, 7.5, 8, 8.5, 9, 9.5, 10, 10.5, 11, 11.5  & Pre-existing gas mass and inflow mass combined\\
SFH & average.sfe, fast.sfe, slow.sfe  & $\rm{SFE}_0=10^{-9}\ \rm{yr}^{-1}$, $\rm{SFE}_0=10^{-8.5}\ \rm{yr}^{-1}$, $\rm{SFE}_0=10^{-9.5}\ \rm{yr}^{-1}$\\
 & average\_bursts.sfe, fast\_bursts.sfe, slow\_bursts.sfe  & Star formation bursts added\\
  \hline
IMF & Chab, TopChab,  & \citet{Chabrier2003}, top-heavy Chabrier (Eqn. \ref{topchab}), \\
 & Salp, Kroup & \citet{Salpeter1955}, `Galactic-field' \citet{Kroupa2003}\\
$f_{\rm recy}$ & 0.1, 0.25, 0.5, 1, 2  & Recycling timescale factor: Eqn. \ref{recycle2}\\
$y_{\rm SN}$ & tot\_LC18\_R000, tot\_LC18\_R150, tot\_LC\_R300,  & \citet{Limongi2018} with $v_{rot}=0,\ 150,\ 300\ \rm{km/s}$ \\
 & MA92\_ori\_extra, MM02\_000  & \citet{Maeder1992} and \citet{Meynet2002} \\
$y_{\rm AGB}$ & Nugrid, FRUITY, &  \citet{Battino2019},  \citet{Cristallo2015} \\
 & KA18\_low, KA18\_high, & \citet{Karakas2018} (low and high mass loss rates) \\
 &  KA10,  VG97 & \citet{Karakas2010}, \citet{vandenHoek1997} \\ 
 \hline
$SN_{\rm red}$ & 1, 5, 20, 80  & Factor SN dust yields are reduced by\\
$M_{\rm destr}$ & 0, 15, 30 $M_\odot$ & Destruction by SN shocks: Eqn. \ref{destreqn} \\
$k_{\rm frag}$ &  0.0, 0.05, 0.5, 1, 5  & photo-fragmentation of dust grains: Eqn. \ref{frageqn}\\
$k_{\rm gg,cloud}$ & 1000, 2000, 4000, 8000, 16000  & Cloud grain growth: Eqn. \ref{ggcloudeqn}\\
$k_{\rm gg,dif}$ & 0, 5, 10 $\rm Gyr^{-1}$  & Diffuse grain growth: Eqn. \ref{ggcloudeqn}\\
$f_{\rm dif}$ & 0.2, 0.3, 0.4  & Maximum dust-to-metal ratio in diffuse ISM: Eqn. \ref{ggcloudeqn} \\ 
 \hline \hline
\end{tabular}
\label{gridtable}
\end{table*}

\begin{itemize}
\item \textbf{Ten} different total galaxy masses (sum of pre-existing initial gas mass and integrated mass of pristine inflows). These masses are logarithmically spaced between $M_{\rm tot}=10^7\,M_{\odot}$ and $M_{\rm tot}=10^{11.5}\,M_{\odot}$.  
\item \textbf{Six} SFHs: bursty and non-bursty SFHs for each of the three reference SFE ($\rm{SFE}_0=10^{-8.5}\ \rm{yr}^{-1}$ (fast), $\rm{SFE}_0=10^{-9}\ \rm{yr}^{-1}$ (average), and $\rm{SFE}_0=10^{-9.5}\ \rm{yr}^{-1}$ (slow).
\item \textbf{Four} different IMFs $\phi(m)$: We run models with the \citet{Chabrier2003} IMF, the \citet{Salpeter1955} IMF, the `Galactic-field' \citet{Kroupa2003} IMF (as opposed to the standard \citet{Kroupa2001} IMF, which is more similar to the \citet{Chabrier2003} IMF) and a top-heavy Chabrier IMF described by: 
\begin{align}
\phi(m)= 
\begin{cases}
     \dfrac{0.402 e^{(-(\log(m)+1.102)^2.)}}{m} & \text{if } m\leq 1\\
     \\
     0.108 m^{-1.8}             & \text{otherwise},
\end{cases}
\label{topchab}
\end{align}
where m is the stellar mass in solar masses.
\item \textbf{Five} metal yield tables for SNe ($mp_{Z,\rm SN}$): We use various tables from the literature to implement the amount of metals expelled by SNe. Our first three tables are the yield tables from Limongi and Chieffi (2018, hereafter LC18) using their recommended set of yields (which is based on the mixing and fallback scheme and produces black holes for $m>25 M_\odot$) and rotation velocities of $v_{\rm rot}=0\ \rm{km/s}$, $v_{\rm rot}=150\ \rm{km/s}$ and $v_{\rm rot}=300\ \rm{km/s}$ respectively. We have also implemented the yields from \citet{Maeder1992} and \citet{Meynet2002}. 
\item \textbf{Six} metal yield tables for AGB stars ($mp_{Z,\rm AGB}$):  We use yield tables from \citet{vandenHoek1997}, \citet{Karakas2010}, the NuGrid collaboration \citep[e.g.][]{Ritter2018,Battino2019}, the FRUITY collaboration \citep[e.g.][]{Cristallo2015} and two tables from \citet{Karakas2018} corresponding the high and low mass loss rates respectively.
\item \textbf{Five} scaling factors for the outflow recycling timescale ($f_{\rm recy}$). 
\item \textbf{Four} SN dust yields tables $y_d(m)$, which are taken from \citet{Todini2001} and scaled down by respectively a factor $\rm SN_{\rm red}$ of 1, 5, 20 and 80.
\item \textbf{Four} values for the amount of dust mass destroyed per SN for a MW-like galaxy $M_{\rm destr}$.
\item \textbf{Five} values for the photo-fragmentation parameter $k_{\rm frag}$.
\item \textbf{Five} values for the cloud dust grain growth scaling factor $k_{\rm gg,cloud}$.
\item \textbf{Three} values for the diffuse dust grain growth scaling factor $k_{\rm gg,dif}$.
\item \textbf{Three} values for the fraction of metals that are available for grain growth in the diffuse ISM $f_{\rm dif}$. From these values we have also determined the values for the fraction of metals that are available for grain growth in clouds $f_{\rm cloud}$ as $f_{\rm cloud}$=$f_{\rm dif}\times2.45$ following the standard ISM THEMIS dust prescription (Jones et al. {\it in preparation}).
\end{itemize}

There are a large number of free parameters in our chemical evolution model. If we were to vary all of these parameters in one large grid, it would be necessary to run more than 97 million models. Unfortunately this is not computationally feasable, even with the relatively simple and computationally light chemical evolution model we employ. Therefore, we take another approach and split the free parameters in two groups. The first group affects metal and star formation related parameters. The second group of free parameters affects only the dust content of the simulated galaxy and can thus be decoupled from the metal and star formation related parameters. 

We thus first vary only the free parameters that affect the metal properties (first six bullet points above) and ignore dust for now, which results in a grid of 36,000 models (Table \ref{gridtable}). We put statistical constrains on which models are most likely by comparing them to the observed properties (excluding dust) of our nearby galaxy samples. The results are presented in Sections \ref{visZgrid} and \ref{MCMCresults_Z}. Next we use the best fitting metal-parameters that come out of this analysis, and vary the model parameters that affect the galaxy dust mass, while keeping the metal parameters constant. The total galaxy masses and SFH are again varied here, as even with a fixed prescription for building metals, galaxies of different masses and at different evolutionary states are still necessary. The dust parameter grid consists of 162,000 models and is presented in Table \ref{gridtable}. This grid of dust parameters is illustrative (see Section \ref{visMdgrid} for the parameter dependencies). For the statistical results in Section \ref{MCMCresults_Md}, we use a direct (continuous) parameter search for the dust parameters. 

\section{Statistical framework}
\label{MCMC}
We have developed a statistical framework to compare our chemical evolution models described in the previous section to the observed galaxy properties in Section \ref{sec:observations}. In order to get constraints on the model parameters, we use a Bayesian Monte Carlo Markov Chain (MCMC) approach. Here we start with the metal grid, we simultaneously compare the observed gas mass, stellar mass, SFR, 12+log(O/H) and log(N/O) to the model predictions for those parameters for the chemical evolution models in the metal grid described in the previous section. For the dust grid we add dust masses to the above observations and compare to models where the dust parameters are varied continuously, as described at the end of this section.

The likelihood function for our models is obtained by combining the likelihood of the individual galaxies:
\begin{align}
   \begin{split}
\mathcal{L}(\boldsymbol{x_{\rm obs}}\ &|\ \textrm{Model}_{(\textrm{IMF},mp_{Z,\textrm{SN}},mp_{Z,\textrm{AGB}},f_{\rm recy},M_{\rm tot},\textrm{SFH})})=\\
 \prod^i \mathcal{L}(\boldsymbol{x_{\rm obs,i}}\ &|\ \textrm{Model}_{(\textrm{IMF},mp_{Z,\textrm{SN}},mp_{Z,\textrm{AGB}},f_{\rm recy},M_{\rm tot},\textrm{SFH})}),
    \end{split}
    \label{Eq:MCMC1}
\end{align}
where $\boldsymbol{x_{\rm obs}}$ are the combined observations of all our galaxies and $\boldsymbol{x_{{\rm obs},i}}$ is a vector with all observed properties for each galaxy $i$. The individual likelihoods are not trivial to determine however for multiple reasons. In most cases these kind of likelihoods are determined to compare an observed and a predicted value of an observable. However in this case we compare an evolutionary track to a single observation. It is a priori not clear which point on the evolutionary track the observed values should be compared to. In order to get around this, we marginalize the likelihood over the different timesteps in the model in order to obtain the total likelihood of at any time.

\begin{align}
   \begin{split}
\mathcal{L}(\boldsymbol{x_{{\rm obs},i}}\ &|\ \textrm{Model}_{(\textrm{IMF},mp_{Z,\textrm{SN}},mp_{Z,\textrm{AGB}},f_{\rm recy},M_{\rm tot},\textrm{SFH})})=\\
\sum^{t} \mathcal{L}(\boldsymbol{x_{{\rm obs},i}}\ &|\ \textrm{Model}_{(\textrm{IMF},mp_{Z,\textrm{SN}},mp_{Z,\textrm{AGB}},f_{\rm recy},M_{\rm tot},\textrm{SFH},t)}).
    \end{split}
        \label{Eq:MCMC2}
\end{align}
In doing this, we account for the fact that the individual observed galaxies did not form at the same time and thus can be at a different point throughout their evolution. In addition, we also know that the individual galaxies did not all evolve from precursors of the same total galaxy mass and that they could have had quite different SFHs. To account for this, we marginalize over all the $M_{tot}$ and SFH options in the grid and determine the total likelihood for any combination of IMF, $mp_{Z,\textrm{SN}}$, $mp_{Z,\textrm{AGB}}$ and $f_{\rm recy}$. In other words, we study ‘families’ of models (represented by the vector $\boldsymbol{\rm{Models}}$) with same chemical evolution parameters, yet where the total galaxy mass, SFH and formation time are allowed to vary.

\begin{align}
   \begin{split}
 & \mathcal{L}(\boldsymbol{x_{{\rm obs},i}}\ |\ \boldsymbol{\rm{Models}}_{(\textrm{IMF},mp_{Z,\textrm{SN}},mp_{Z,\textrm{AGB}},f_{\rm recy})})=\\
 \sum^{M_{\rm tot}} \sum^{\textrm{SFH}}& \mathcal{L}(\boldsymbol{x_{{\rm obs},i}}\ | \\
 & \textrm{Model}_{(\textrm{IMF},mp_{Z,\textrm{SN}},mp_{Z,\textrm{AGB}},f_{\rm recy},M_{\rm tot},\textrm{SFH})})=\\
  \sum^{M_{\rm tot}} \sum^{\textrm{SFH}} & \sum^{t} \mathcal{L}(\boldsymbol{x_{{\rm obs},i}}\ | \\
  & \textrm{Model}_{(\textrm{IMF},mp_{Z,\textrm{SN}},mp_{Z,\textrm{AGB}},f_{\rm recy},M_{\rm tot},\textrm{SFH},t)}).
      \end{split} 
    \label{Eq:MCMC3}
\end{align}

Here it has to be kept in mind that we are simultaneously fitting all observables (gas mass, stellar mass, SFR, 12+log(O/H), log(N/O) and dust mass). The likelihood for a given timestep in one of the models is thus:

\begin{align}
   \begin{split}
\mathcal{L}(\boldsymbol{x_{{\rm obs},i}}\ |\ \textrm{Model}_{(\textrm{IMF},mp_{Z,\textrm{SN}},mp_{Z,\textrm{AGB}},f_{\rm recy},M_{\rm tot},\textrm{SFH},t)})=\\
\prod^{x} \left( \dfrac{1}{\sqrt{2\pi\sigma_{x_i}^2}} \times e^{\left( \dfrac{- (x_{i,{\rm obs}}-x_{\rm model})^2}{2\ (\sigma_{x_i}^2+\sigma_{\rm model}^2)}       \right)} \right), 
    \end{split}
    \label{Eq:MCMC4}
\end{align}
where $x_{i,\rm obs}$ are the observables for each observation $i$, and $x_{\rm model}$ the observables for a given model (at a given time t). $\sigma_{x_i}$ and $\sigma_{\rm model}$ are the uncertainties on the observations and the model uncertainty. Here the model uncertainty is included to account for systematic errors in model conversion factors as well as the limited sampling we can do due to the discrete nature of our models (especially for bursty models, where the bursts themselves have a very limited sampling).  For the stellar mass, gas mass, dust mass and log(N/O) we use $\sigma_{\rm model}=0.3~\rm{dex}$, which roughly corresponds to the intrinsic scatter in the observed sample. For 12+log(O/H), the observed scatter is smaller and we use $\sigma_{\rm model}= 0.15\ \rm{dex}$. For the SFR, we use a larger amount of scatter as there is both larger intrinsic scatter in the observations and there is an inherent variation due to the bursts. The interburst periods and burst periods are randomly generated, so if the same model is generated multiple times there will be variation in the resulting SFR. Therefore we use $\sigma_{\rm model}= 0.9\ \rm{dex}$ for the SFR. The final likelihood for comparing all available observations to a given family of models is obtained by combining Equations \ref{Eq:MCMC1}-\ref{Eq:MCMC4}: 

\begin{align}
   \begin{split}
\mathcal{L}(\boldsymbol{x_{\rm obs}}\ &|\ \boldsymbol{\rm{Models}}_{(\textrm{IMF},mp_{Z,\textrm{SN}},mp_{Z,\textrm{AGB}},f_{\rm recy})})=\\
 \prod^i \sum^{M_{\rm tot}} \sum^{\textrm{SFH}} \sum^{t} \prod^{x} &  \dfrac{1}{\sqrt{2\pi\sigma_{x_i}^2}} \times {\rm exp}{\left( \dfrac{- (x_{i,\rm obs}-x_{\rm model})^2}{2\ (\sigma_{x_i}^2+\sigma_{\rm model}^2)}       \right)}.
    \end{split}
    \label{Eq:MCMC5}
\end{align}

Using this likelihood function, it is relatively straightforward to fit the chemical evolution models to the data. 
We determined the posterior probability of our chemical evolution parameters (IMF, $mp_{Z,\textrm{SN}}$, $mp_{Z,\textrm{AGB}}$ and $f_{\rm recy}$) in a Bayesian manner, sampling the posterior Probability Distribution Functions (PDF) using the emcee \citep{Foreman2013} MCMC package for
Python. Effectively, for the metal results we are sampling the indices for the model grid described in Section \ref{sec:modelgrid}. Since the indices of a grid are discrete, we round each number in each proposed MCMC sample before selecting the model\footnote{However for generating the next proposal (the next step in the MCMC chain), the non-rounded values are used as to not introduce any biases.}. The priors are set uniformly (every model is a priori equally likely) and we use 250 walkers and a burn-in phase of 50 ($\times250$) steps followed by 200 ($\times250$) used steps.

For the dust parameters, we have also run a grid of models as described in Section \ref{sec:modelgrid}, which will be used in the following section to illustrate the parameter dependencies. However, for constraining the dust parameters, more realistic results can be obtained  by varying the dust parameters continuously instead of on a discrete grid as for the metal parameters. Unlike the metal parameters, the dust parameters are all numerical values, and can thus be varied continuously. Therefore we generate a MCMC sample where instead of using discrete indices on a grid of models, we generate the parameter values directly and run a chemical evolution model for each proposal in the MCMC chain. The priors on the parameters are uniform for $M_{\rm destr}$, $k_{\rm gg,dif}$, $f_{\rm dif}$ and log-uniform for $\rm SN_{\rm red}$, $k_{\rm frag}$ and $k_{\rm gg,cloud}$. This division has been used because it mirrors the distribution of grid-values for these parameters. The grid-values were in turn distributed that way because they lead to similar spacing between models in Figure \ref{ill_fam_Md} (Section \ref{visMdgrid}). Table \ref{priors} shows the limits of the priors.

\begin{table}
\caption{Priors on the MCMC parameters constrained in this work.}
\centering
\footnotesize
\begin{tabular}{c|cccc} \hline\hline
Parameter & Uniform & Log-uniform & lower & upper \\ 
 &  &  & limit & limit \\ \hline
$\rm SN_{\rm red}$ &  & \checkmark & 1 & 80 \\
$M_{\rm destr}$ & \checkmark &  & 0 $M_\odot$ & 30 $M_\odot$ \\
$k_{\rm frag}$ &  & \checkmark & 0.005 & 5 \\
$k_{\rm gg,cloud}$ &  & \checkmark & 1000 & 16000 \\
$k_{\rm gg,dif}$ & \checkmark &  & 0 $\rm Gyr^{-1}$ & 50 $\rm Gyr^{-1}$\\
$f_{\rm dif}$ & \checkmark &  & 0.1 & 0.4 \\ 
 \hline \hline
\end{tabular}
\label{priors}
\end{table}

The probability for this model is then calculated in exactly the same way as described above. This method is more computationally expensive as large numbers of models need to be run. Therefore we use 100 walkers and a burn-in phase of 10 ($\times100$) steps followed by 100 ($\times100$) used steps. For each of these proposals we run 60 models in order to marginalize our probabilities over the 6 SFH and 10 initial masses (for a total of 600,000 models analysed). We call this framework {\sc BEDE} - Bayesian Estimates of Dust Evolution.

\section{Visualizing parameter dependencies}
\label{sec:vis}
\subsection{Metal parameters}
\label{visZgrid}
\begin{figure}
  \center
\includegraphics[width=\columnwidth]{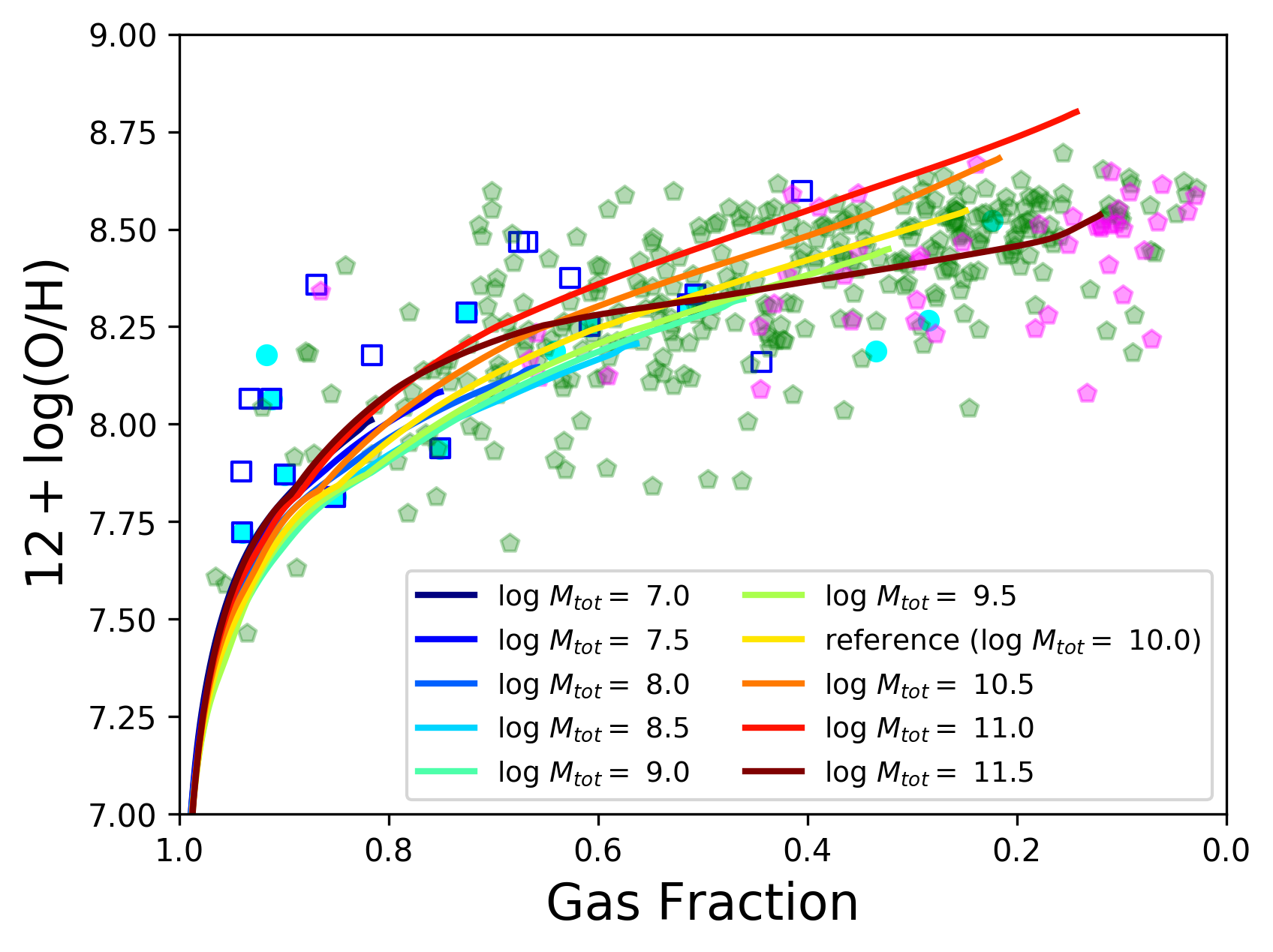}
\includegraphics[width=\columnwidth]{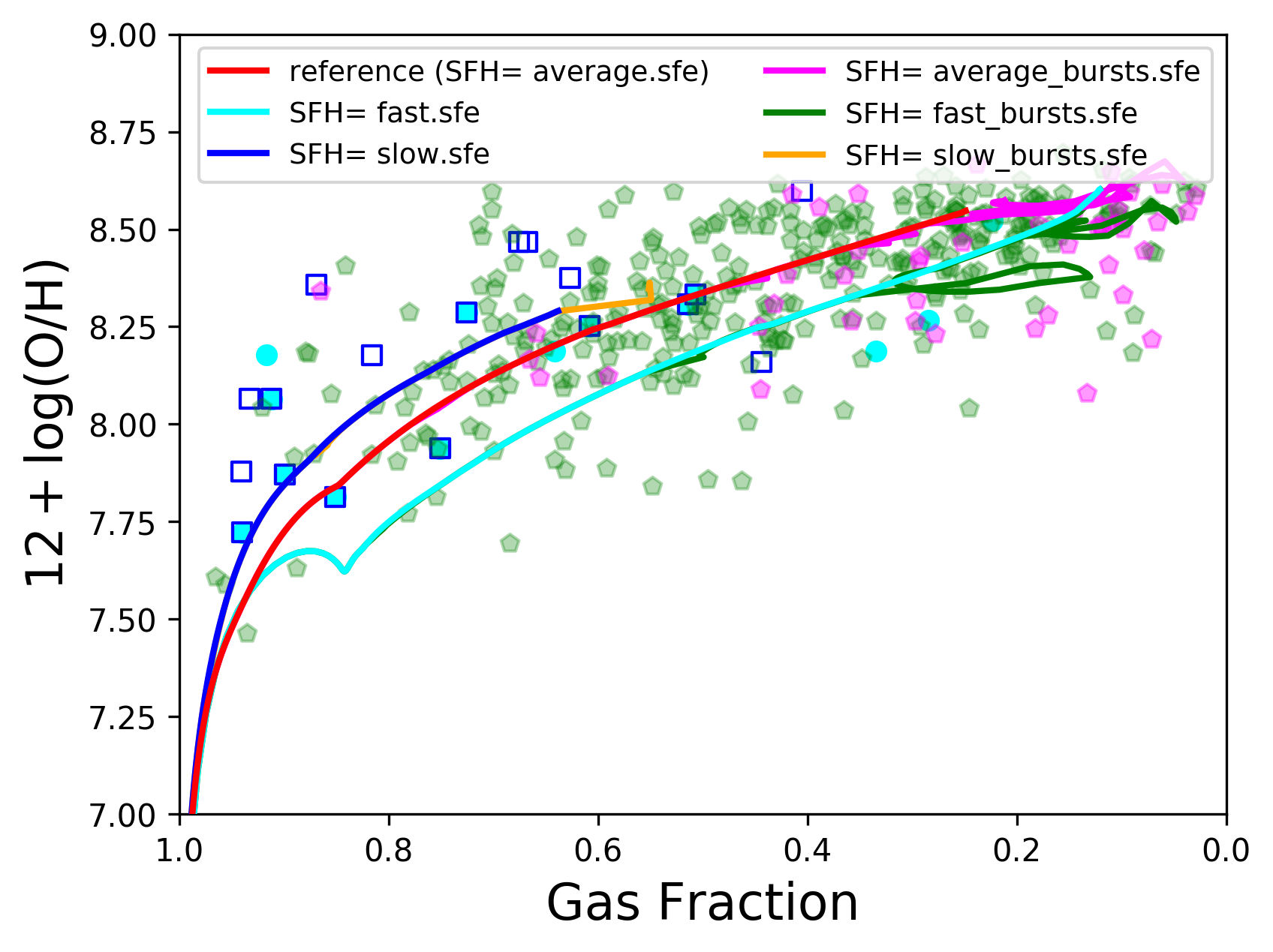}
  \caption{The build up of metallicity with decreasing gas fraction for our observed samples (data points) and one family of chemical evolution models (curves). The data is obtained from the HIGH survey (blue squares), DustPedia (green and purple hexagons for LTG and ETG respectively) and HAPLESS (cyan circles).  \textit{Top:} Changing $M_{\rm tot}$ (sum of initial cloud mass + pristine inflows mass) shows the spread of models in a family (shown by the different colour curves (see legend) with the reference model of $10^{10}\,M_{\odot}$ in yellow). \textit{Bottom:} Changes in the build up of metals due to changing the SFH (with the  reference `average' SFE model shown in red).}
  \label{ill_fam}
\end{figure}

Before we look at the statistical results, it is beneficial to visualize how our chemical evolution models typically depend on their input parameters and how they compare to the nearby galaxy observations. Since the parameter-space is multidimensional and there are a lot of models, it is impossible to compare our models simultaneously in one plot. Therefore we here select one model as a reference model, and vary the chemical evolution parameters one by one to illustrate the main effects they have on the observables. As reference, we will use a model from the family of models with the highest probability from the MCMC fit (detailed in Section \ref{MCMCresults_Z}). Many of the chemical evolution parameters actually affect most or all of the observables, however illustrating all these dependencies would take up too much space in this paper. Here we look at some key plots to illustrate the main variation, with additional focus on the build up of metals. 

\begin{figure*}
  \center
\includegraphics[width=\columnwidth]{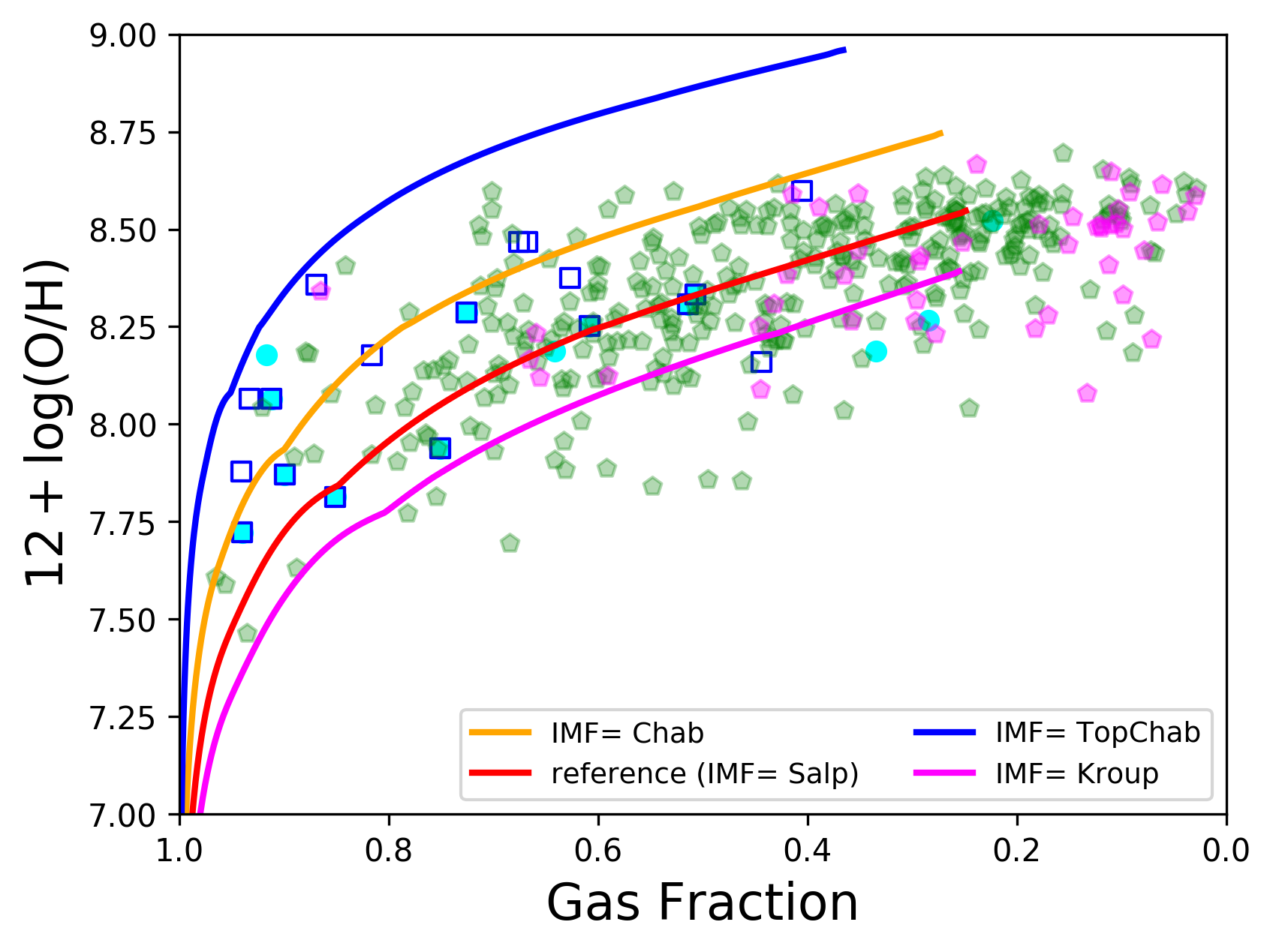}
\includegraphics[width=\columnwidth]{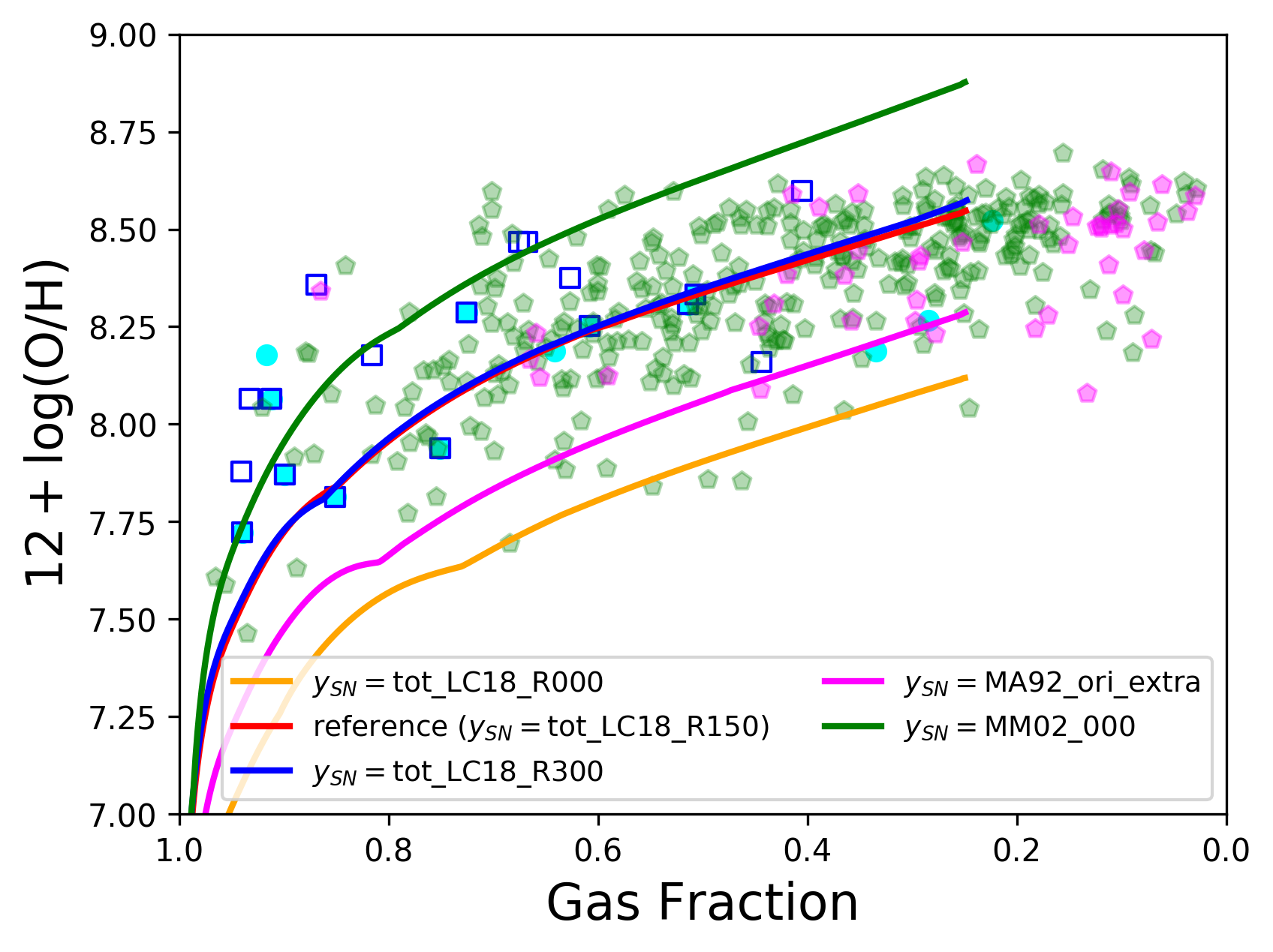}
\includegraphics[width=\columnwidth]{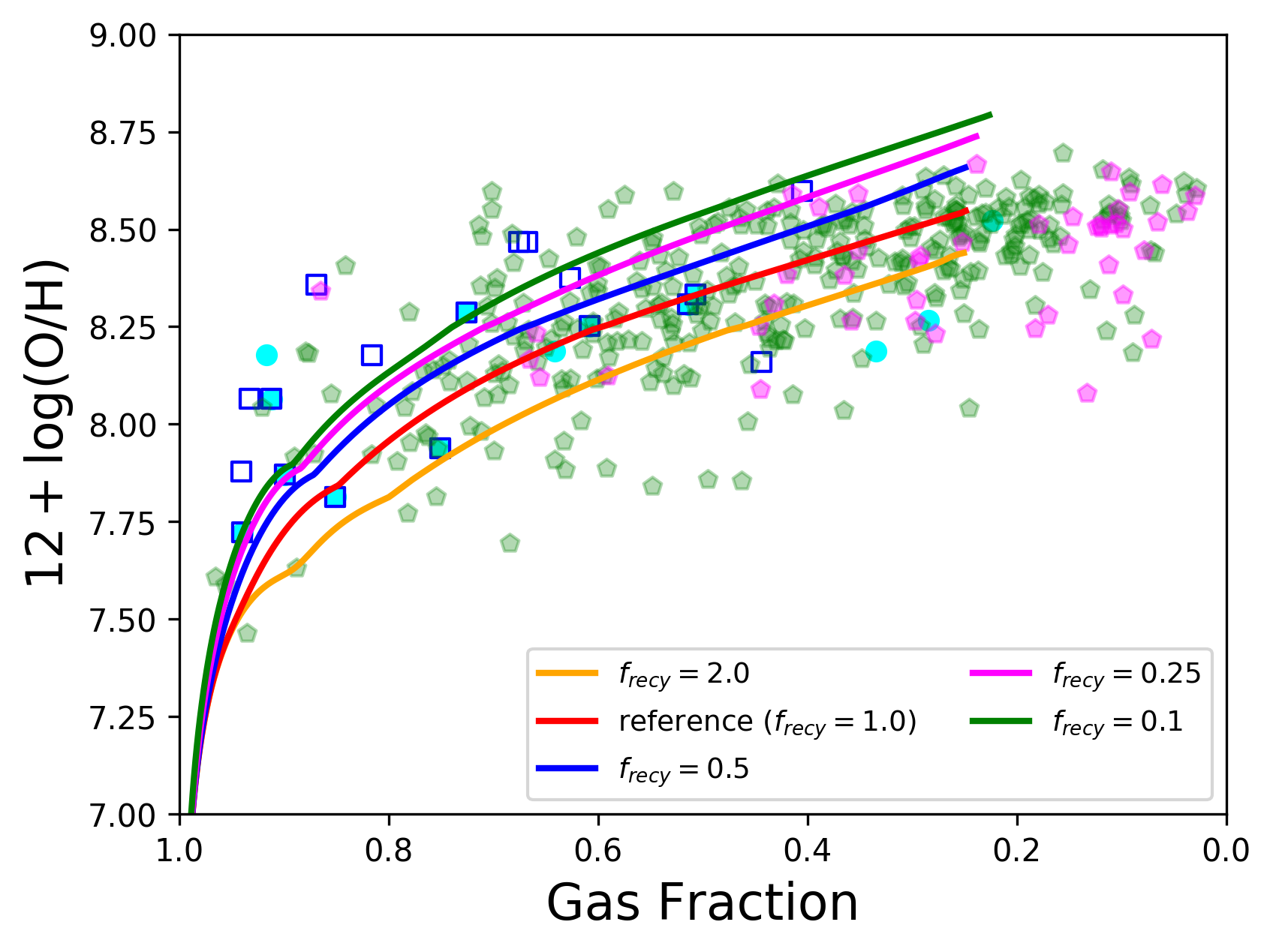}
\includegraphics[width=\columnwidth]{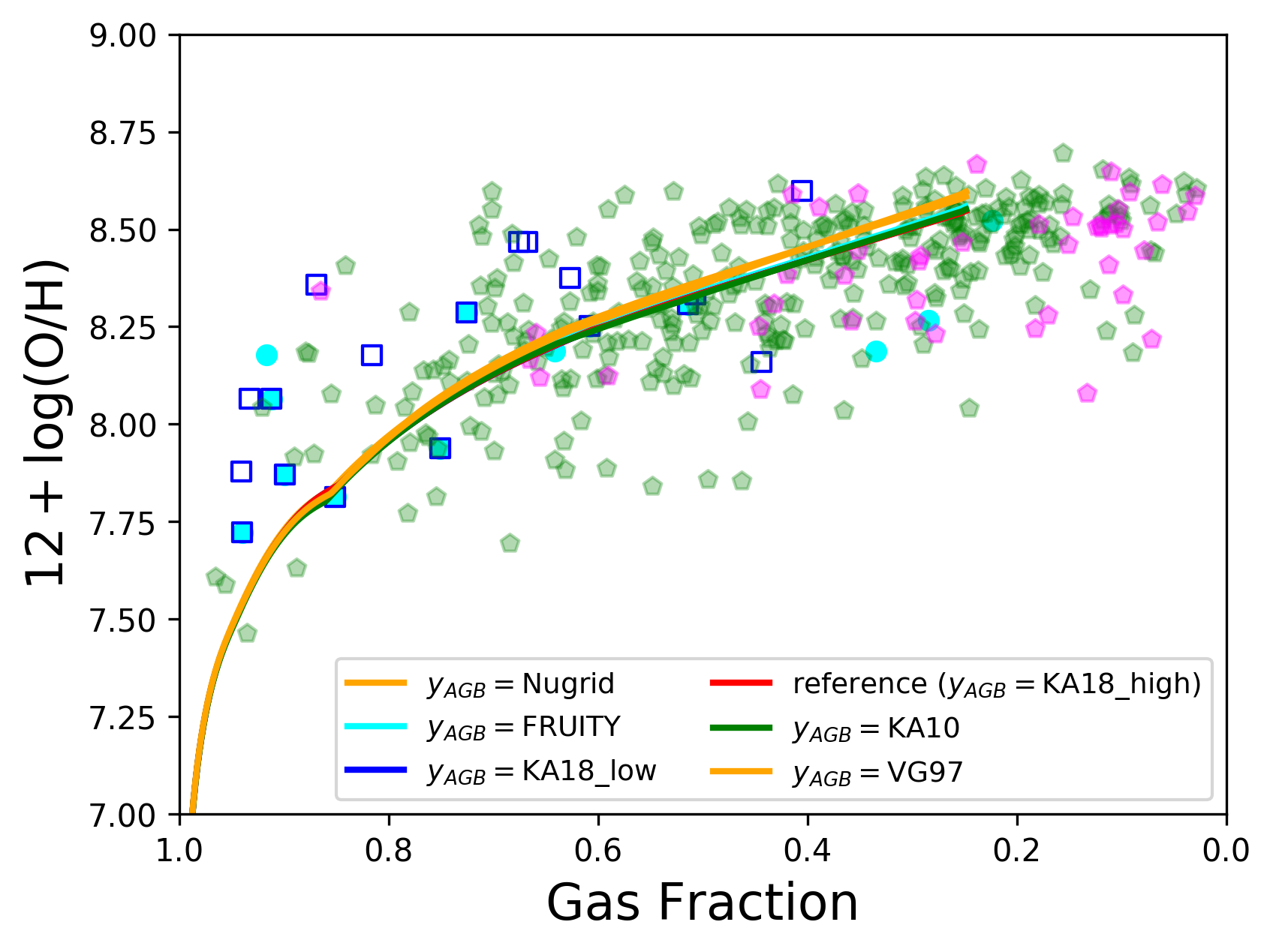}
  \caption{The variation in the chemical evolution models due to changing the IMF (\textit{top left}), the SN metal yields (\textit{top right}), the recycling scaling factor $f_{\rm recy}$ (\textit{bottom left}) and the AGB metal yields (\textit{bottom right}). Symbols for the observed datapoints are the same as in Figure \ref{ill_fam}. The red curve shows the reference model in each subplot, see legend for more details on the parameter values used.}
  \label{ill}
\end{figure*}

Using the MCMC trace described in Section \ref{MCMC}, we determine the family of models with the highest probability as the one with a Salpeter IMF, standard recycling of outflows ($f_{\rm recy}=1$), LC18 ($v_{rot}$ = 150 km/s) SN metal yields and \citet{Karakas2018} high mass-loss AGB metal yields (see Section \ref{MCMCresults_Z}). Here we remind the reader that families of models have the same chemical evolution parameters, but the total galaxy mass, SFH and formation time are allowed to vary. The observed data are compared to all models in the family, and thus the general location and spread of the whole family of models is more important than the location of an individual model. In Figure \ref{ill_fam} (\textit{top}), we illustrate this spread and show the build up of metals (12+log(O/H)) as gas is converted into stars for models with different $M_{\rm tot}$ that are part of this family. The gas fraction of the galaxy is defined as $f_g = M_g/(M_g+M_s)$. Each point along the plotted evolutionary track for each $M_{\rm tot}$ correspond to a different time in its evolution (or equivalently to the state of the galaxy at the current age of the universe for a different formation time). Each galaxy is compared to each point on this track, as we do not know when the galaxy was formed. Similarly, Figure \ref{ill_fam} (\textit{bottom}) shows models of this same family with different SFH and $M_{\rm tot}=10^{11} M_\odot$. The observed nearby galaxies do not need to all have the same $M_{\rm tot}$, SFH or formation time. As the reference model in subsequent plots, we use the model of this family that has $M_{\rm tot}=10^{11} M_\odot$ and the non-bursty average (${\rm SFE}_0=10^{-9}\ yr^{-1}$) SFH. The red line in all panels of Figure \ref{ill_fam} and Figure \ref{ill} shows the same model.

In Figure \ref{ill}, we vary the remaining chemical evolution parameters one by one, and again show the build up of metals with decreasing gas fraction. Here each line belongs to a different family of models. These are the metal parameters we want to constrain in this work (Section \ref{MCMCresults_Z}). The top left panel shows how the IMF affects the build up of metals. We find considerable differences in how much metals are released by each IMF. Massive stars expel relatively more metals and evolve faster. IMFs with a larger fraction of massive stars will thus produce more metals. We indeed see that the top-heavy Chabrier IMF described in Section \ref{sec:modelgrid} produces most metals, followed by the Chabrier IMF, the Salpeter IMF and finally the Kroupa (2003) IMF. The top right panel of Figure \ref{ill} shows the how the build up of metals is affected by the SN yield tables used. Theoretical calculations of the amount of metals expelled in a SN event differs among authors, with the yield tables of \citet{Meynet2002} having the largest amount of expelled metals, followed by the rotating (300 and 150 km/s respectively) SN models of LC18. The yields from \citet{Maeder1992} are next and finally the non-rotating models from LC18. Note that here we are not varying multiple chemical evolution parameter simultaneously and are not showing all observables.  We thus cannot make much inference about which IMF or yields are best by comparing the models to the data in only this one plot. Even though the non-rotating LC18 yields result to a poor fit to the data in this plot, they result in much a better fit if they are e.g. combined with a top-heavy IMF. We explore the full metal parameter space in Section \ref{MCMCresults_Z}.

The bottom left panel shows how the recyling time scaling factor $f_{\rm recy}$ (see Section \ref{inout}) affects the slope of the build up of metals. When $f_{\rm recy}$ is small, outflows are recycled very fast. As a result, there are not as many metals suspended in the IGM, and the build up of metals within the galaxy is faster. Vice versa, a large $f_{\rm recy}$ leads to a slow build up of the galaxy's metallicity as a lot of metals are in the IGM. Changing $f_{\rm recy}$ gives us a way to scale the amplitude of the effect the outflows have on the evolution of galaxies. 

\begin{figure}
  \center
\includegraphics[width=\columnwidth]{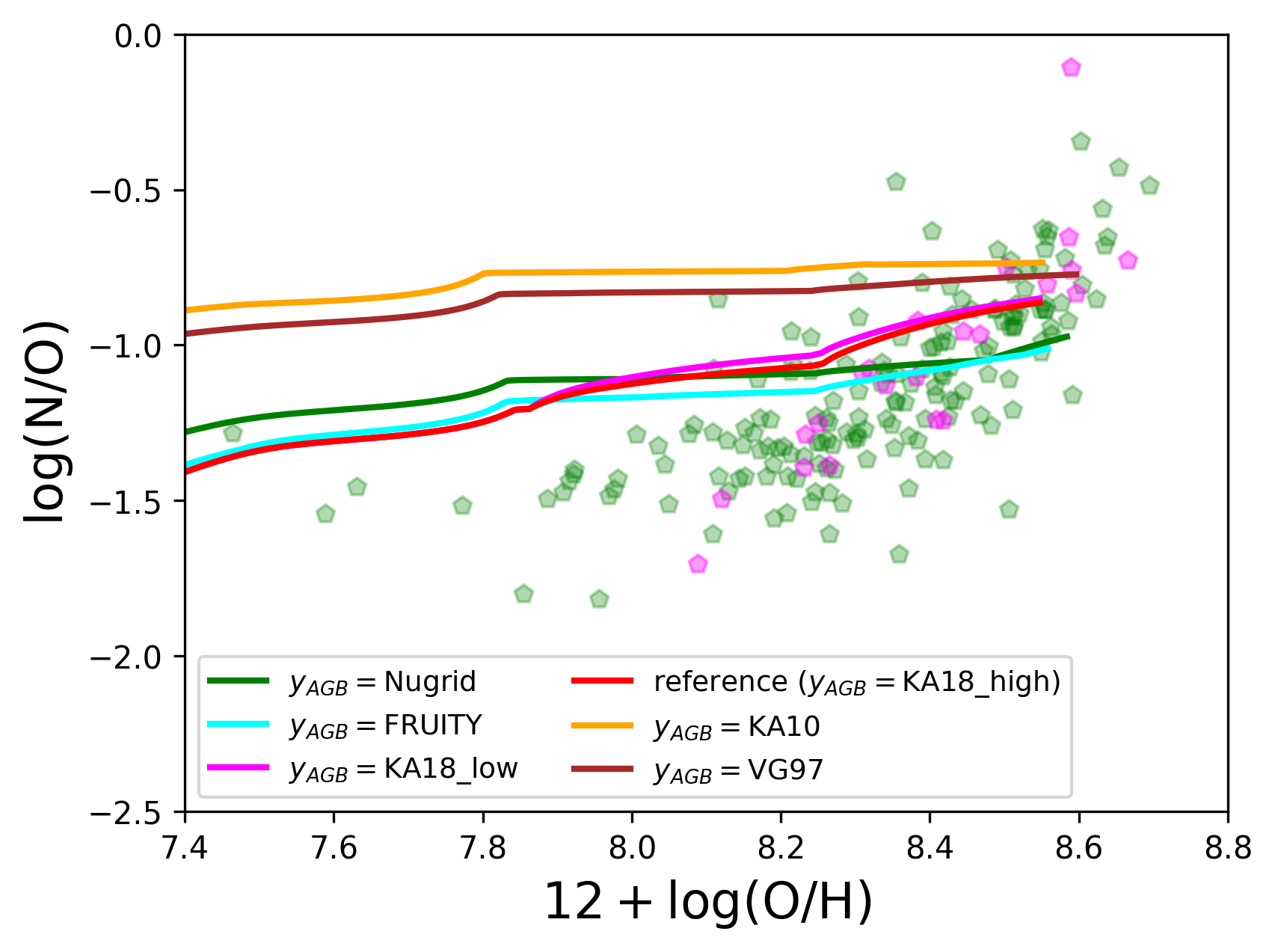}
  \caption{The variation of the Nitrogen-to-Oxygen ratio with metallicity for the different input AGB metal yields (Table~\ref{gridtable}). Changing the input AGB metal yields changes the N/O ratio in spite of not changing the metallicity (see also Figure \ref{ill}). Symbols for the observed datapoints are the same as in Figure \ref{ill_fam}. }
  \label{AGB}
\end{figure}

In the bottom right panel of Figure \ref{ill} we see that the build up of oxygen (or the total mass of metals) is little affected by which AGB metal yields are used. The vast majority of metals is expelled by SN. This does not mean however that AGB yields are unimportant. In Figure \ref{AGB} we show how the Nitrogen-to-Oxygen ratio changes with metallicity for models with different AGB yields. Here it is clear that the predicted Nitrogen-to-Oxygen ratio is significantly affected by choice of AGB yields from the literature. Older AGB yield tables such as the ones from \citet{vandenHoek1997} or \citet{Karakas2010} produce too high Nitrogen-to-Oxygen ratios at low metallicities. The other yields used provide a better match, with the \citet{Karakas2018} yields with high mass loss rates providing the best match to the build up of log(N/O) with increasing metallicity.  	    

\begin{figure}
  \center
\includegraphics[width=\columnwidth]{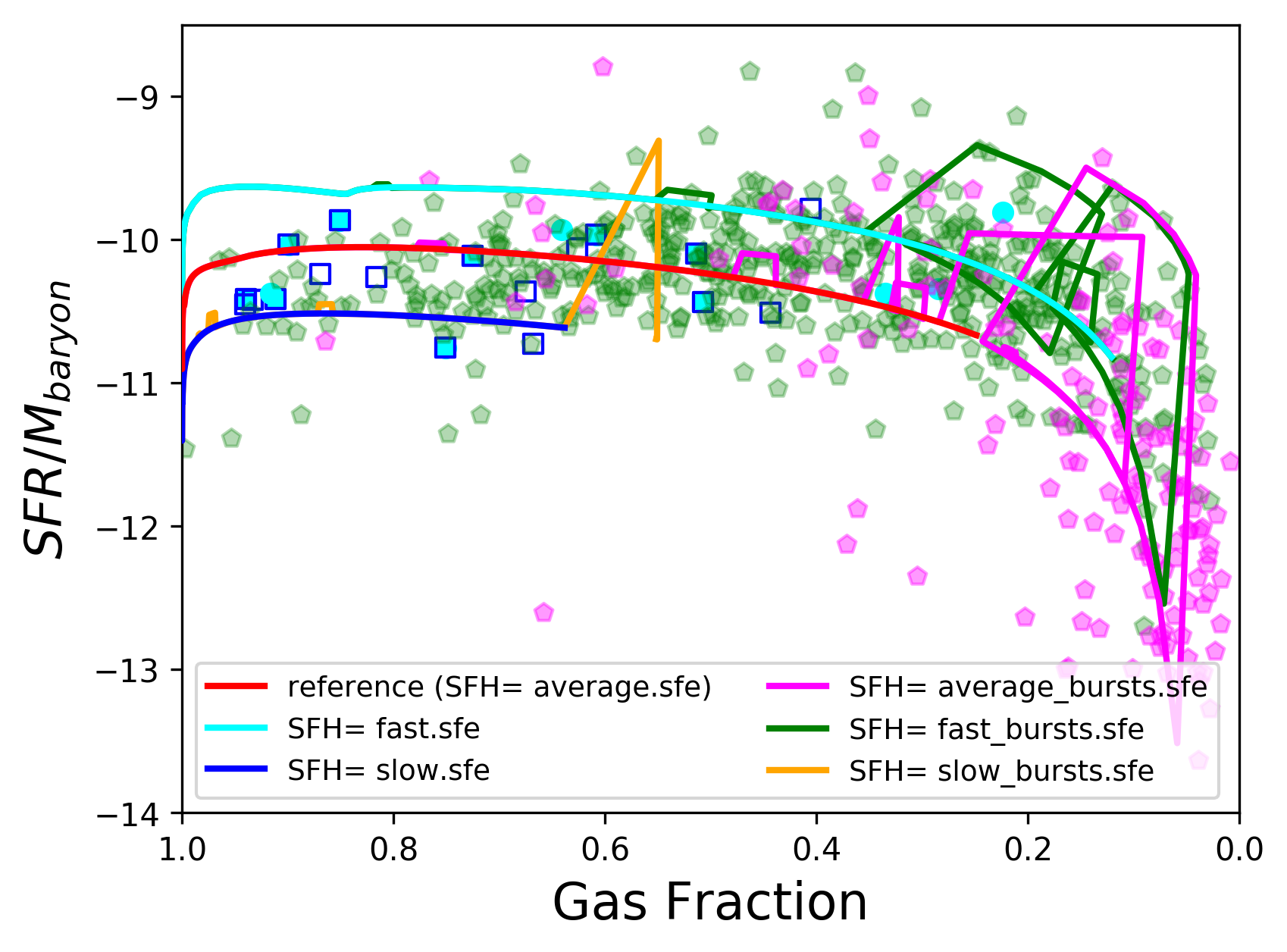}
  \caption{An illustration of the various kind of SFHs used in this work. Note that the bursty and non-bursty models overlap during the inter-bursts periods. Symbols for the observed datapoints are the same as in Figure \ref{ill_fam}. Different coloured curves illustrate different star formation efficiencies.}
  \label{SFR}
\end{figure}

\begin{figure*}
  \center
\includegraphics[width=\columnwidth]{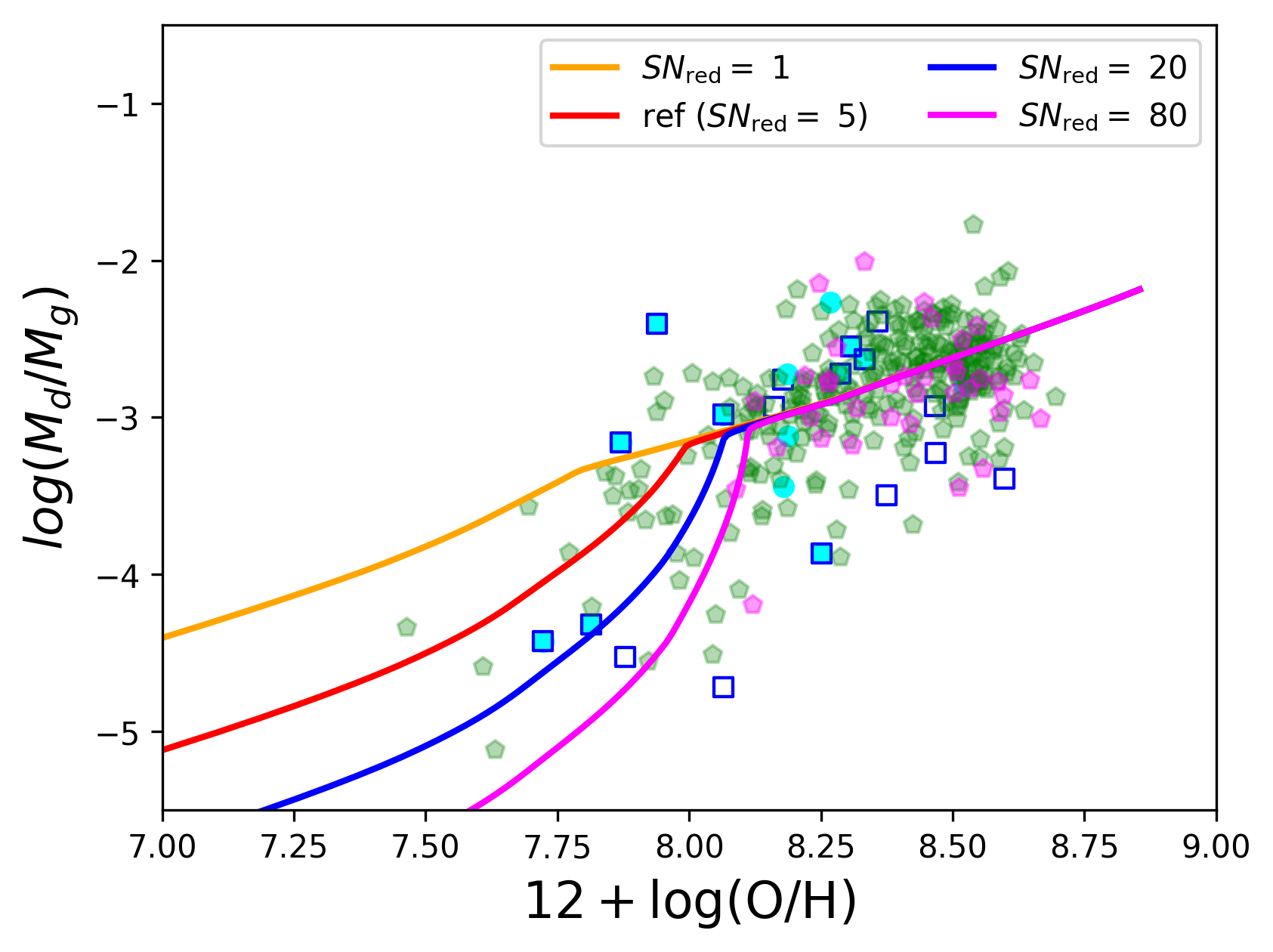}
\includegraphics[width=\columnwidth]{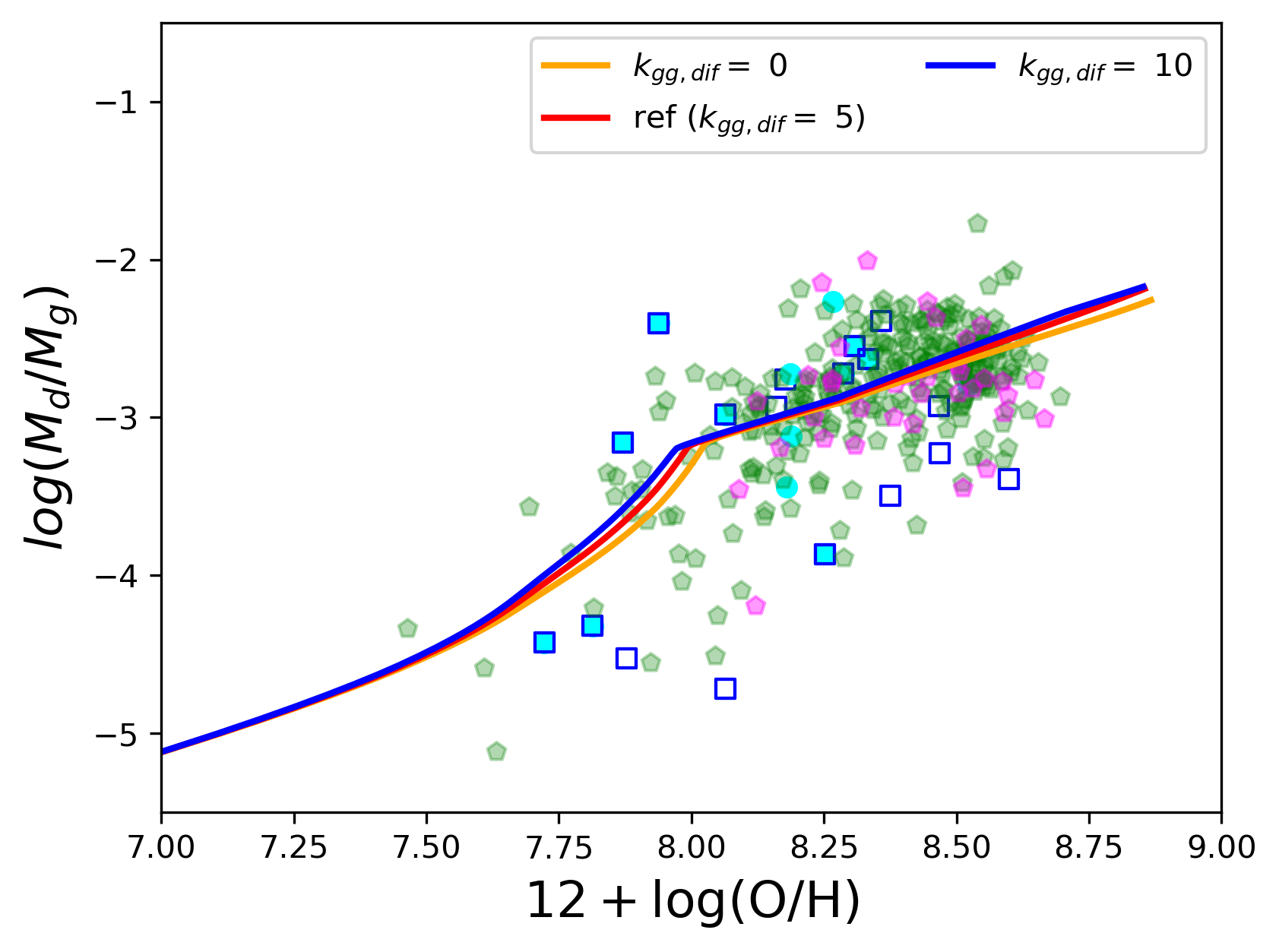}
\includegraphics[width=\columnwidth]{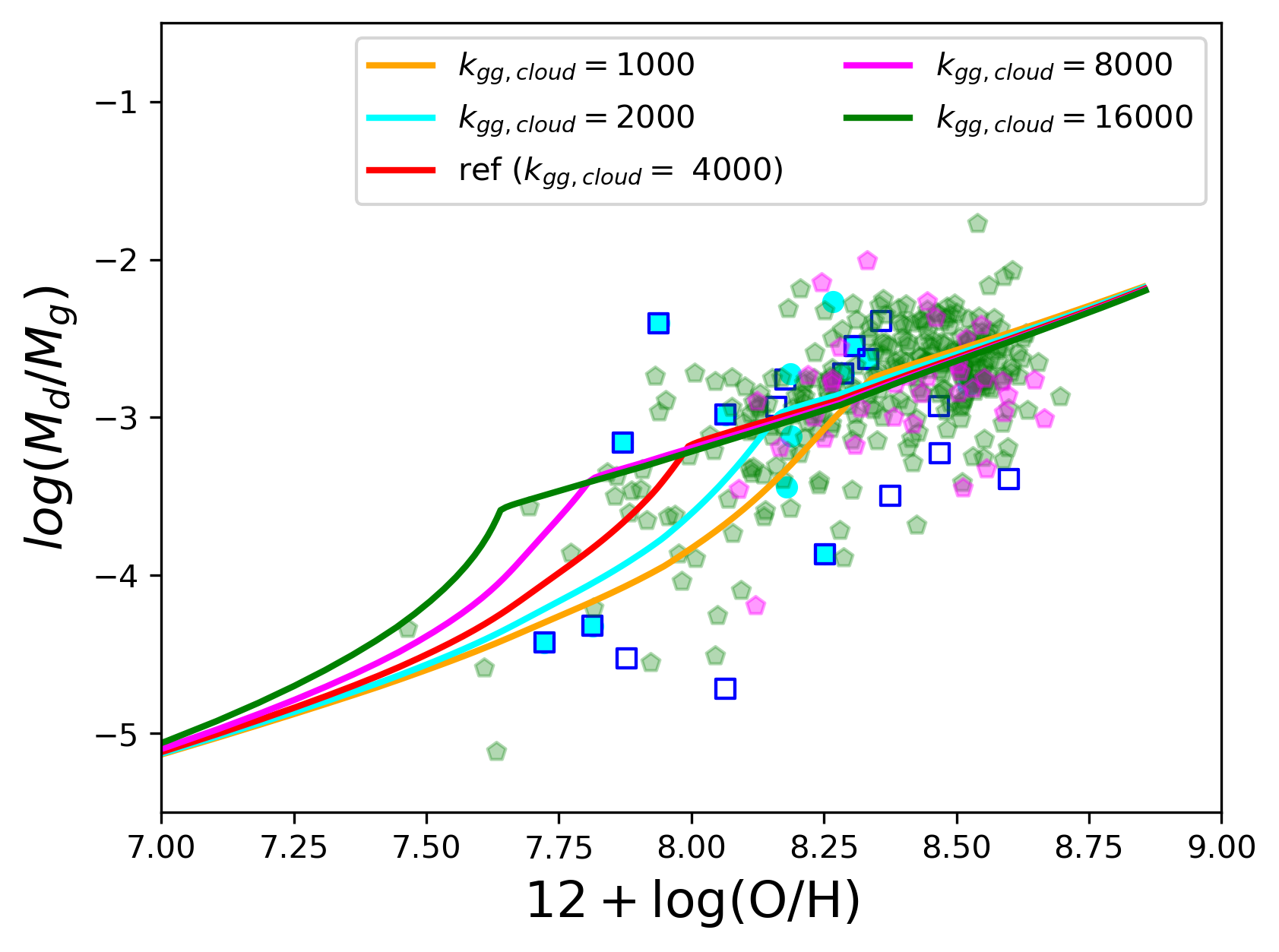}
\includegraphics[width=\columnwidth]{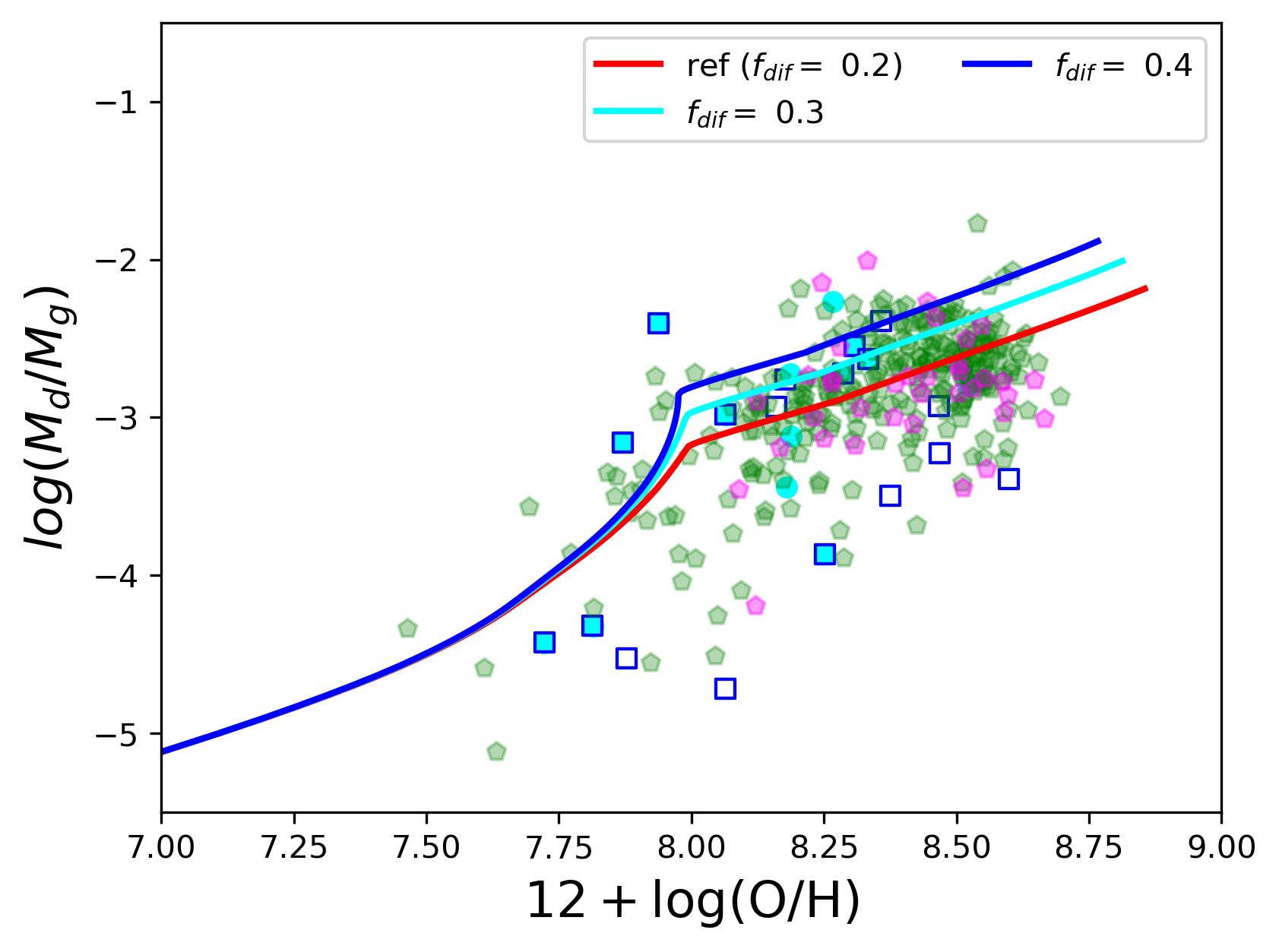}
\includegraphics[width=\columnwidth]{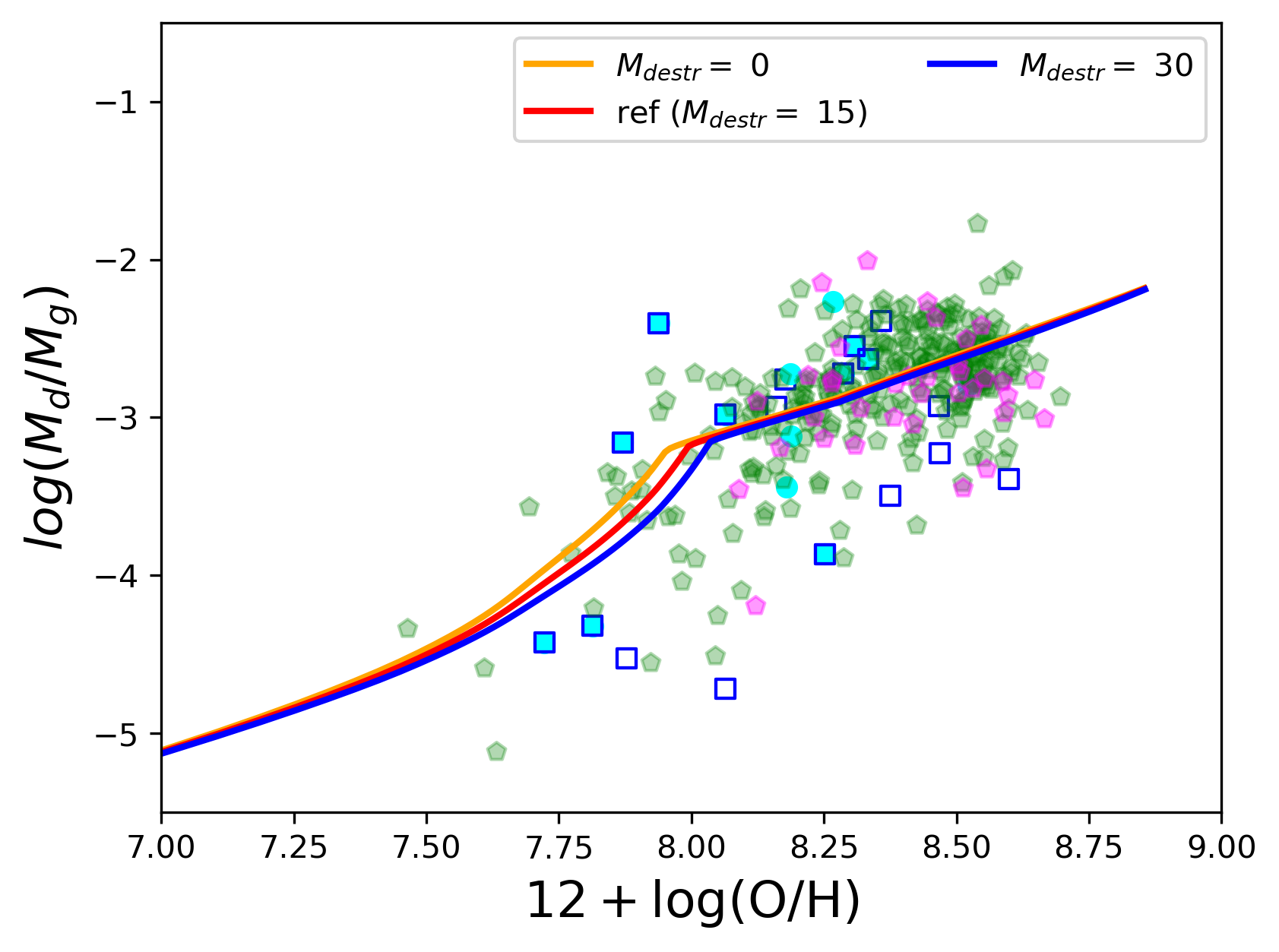}
\includegraphics[width=\columnwidth]{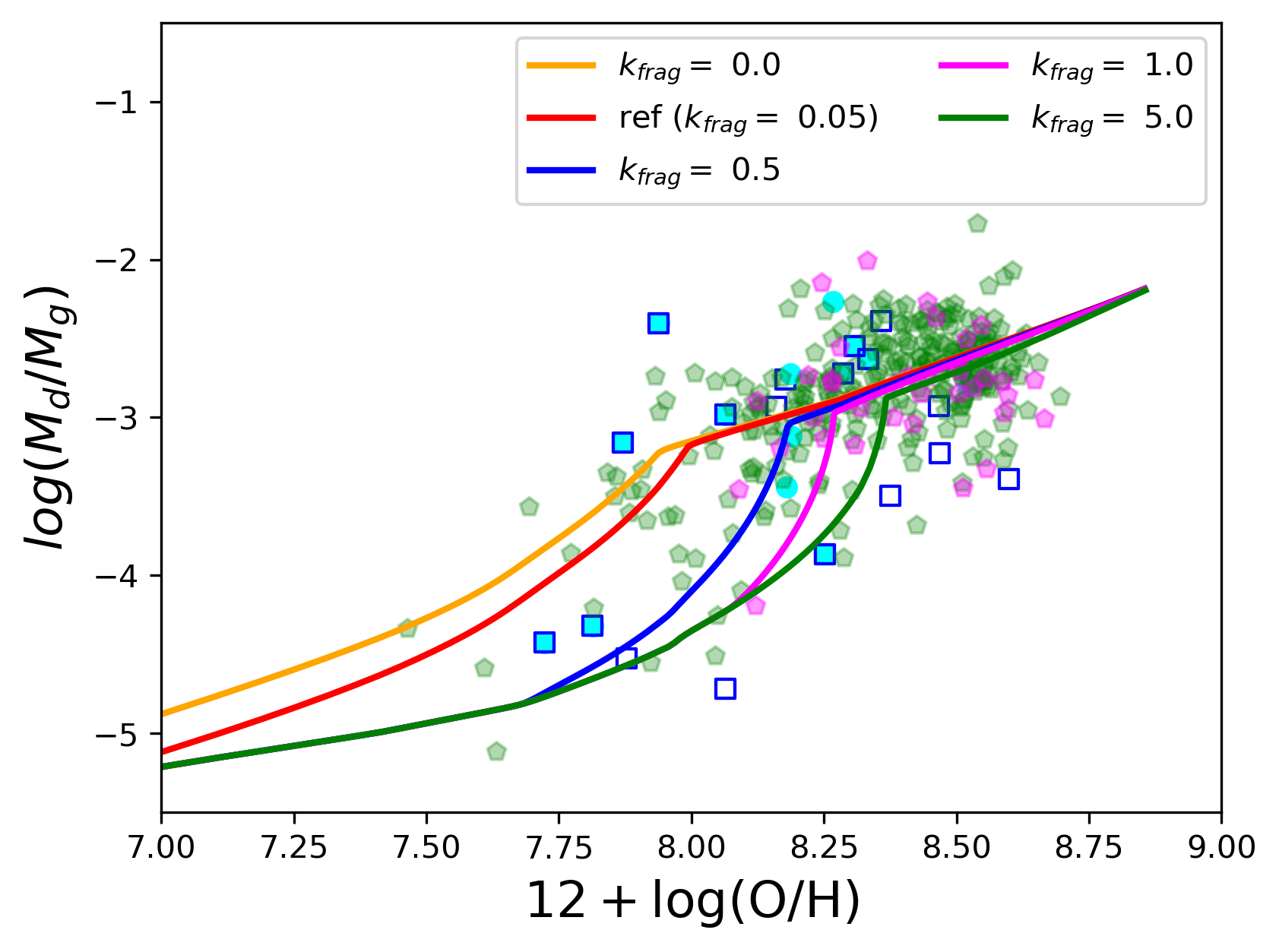}
  \caption{The build up of the dust-to-gas mass ratio with metallicity for our observed samples (data points, see Figure \ref{ill_fam}) and models (coloured curves) illustrating the variation in the chemical evolution models due to changing the SN dust yields (\textit{top left}), the diffuse grain growth rate (\textit{top right}), the cloud grain growth rate (\textit{middle left}), the fraction of metals available for grain growth (\textit{middle right}), the mass of dust destroyed by SN shocks for a MW-type galaxy (\textit{bottom left}), and the dust photo-fragmentation rate (\textit{bottom right}). The reference model for each parameter  is shown in red, see legend for more details on the parameter values used.}
  \label{ill_fam_Md}
\end{figure*}

Given that our SFR prescription is empirical, we show that the resulting values are sensible in Figure \ref{SFR}. Here we plot the SFR/$M_{\rm baryon}$ as this normalisation allows us to more easily compare massive and dwarf galaxies in terms of the SFR relative to their size. The different SFH together span the parameter space quite well\footnote{There are some outliers in Figure \ref{SFR} with much lower SFR/$M_{\rm baryon}$ that are not reached by our models. However these outlying sources are ETG's which often have quite poorly determined SFR and molecular gas masses. The outlying sources also have quite large uncertainties and can thus safely be ignored.}. Note that the SF bursts are introduced randomly, so each bursty model will have its burst at different gas fractions. The various models thus together span the SFR parameter-space excellently. For each combination of chemical evolution parameters, we see there are six different SFHs based on three different reference ${\rm SFE}_0$ (See Section \ref{sec:modelgrid}). For each of these, there is one SFH where bursts have been included, and one SFH without bursts. During the bursts, the SFR are increased for a given number of timesteps as discussed above. Because the mass loading factors do not change during the bursts, the increased SFR also infer significant outflows. The SFR and outflows together significantly reduce the gas fraction during the burst. However after the bursts, a significant fraction of the outflows will be recycled in a short time-span, especially for more massive galaxies (with lower gas fractions). This results in the increase in gas fraction that can be seen after the bursts at lower gas fractions. 

\subsection{Dust parameters}
\label{visMdgrid}

We illustrate the variation introduced by the different dust parameters by plotting how the dust-to-gas mass ratio evolves with increasing metallicity for both the models and observed data. We choose the model with ${\rm SN_{ red}}=5$, $M_{\rm destr}=15\ M_{\odot}$, $k_{\rm frag}=0.05$, $k_{\rm gg,cloud}=4000$, $k_{\rm gg,dif}=5$ and $f_{\rm dif}=0.2$ as our reference model. In Figure \ref{ill_fam_Md}, we vary the dust chemical evolution parameters one by one, as compared to this reference model and the data. 

The first panel of Figure \ref{ill_fam_Md} shows how changing the SN dust yield affects the build up of dust as galaxies evolve. When the SN dust yield is reduced by a larger factor ($\rm SN_{red}$), the dust-to-gas ratio is reduced at low metallicities. At higher metallicities the difference is much smaller. This is because at early evolutionary stages, SN dust is the dominant contributor, whereas for more evolved galaxies (once the `critical metallicity' is reached), dust grain growth is dominant. Reducing the SN dust content is a possible avenue to explain the low dust content of galaxies at early evolutionary stages \citep{Remy-Ruyer2014,DeVis2017a,DeVis2019}. 

In the next two panels of Figure \ref{ill_fam_Md} we study the effects of varying the dust grain growth parameters in the diffuse ISM and in clouds. We find the cloud grain growth parameter has a much stronger influence than the diffuse grain growth parameter. For both environments, most variation is introduced at intermediate metallicities. At low metallicities the grain growth is too inefficient since too few metals are available for accretion onto the dust grains. At large metallicities, grain growth is dominant and all of the available metals have accreted onto the dust grains. For each of the illustrated models, the maximum dust-to-metal ratio has been reached for the highest metallicities. The dust grain growth scaling factors $k_{\rm gg,cloud}$ and $k_{\rm gg,dif}$ thus affect when the grain growth becomes dominant (i.e. when the `critical metallicity' is reached) and how fast the metal reservoir accretes onto the dust grains. 

The fourth panel shows how the models change when we change the fraction of metals that are available for grain growth. This mainly determines the maximum dust-to-metal ratio (and thus dust-to-gas ratio at a given metallicity) that is reached at high metallicities. Given the majority of our models reach this maximum dust-to-metal ratio at high $Z$, the $f_{\rm dif}$ parameter is the one that most strongly affects the dust-to-gas ratio at high metallicities. 

The two bottom panels of Figure \ref{ill_fam_Md} show how the different dust destruction mechanisms change the build up of dust. Destruction of grains by SN shocks is not very efficient, the effect is minor even if a single SN destroys 30 $M_{\odot}$ of dust. At low metallicity, the SN shocks are inefficient because the dust-to-gas ratio is very low and so not much dust is present in the gas the shock propagates through. At high metallicities, significant amounts of dust are destroyed by SN, but by this point grain growth has become so efficient that the destroyed dust is rapidly re-accreted onto other dust grains. The photo-fragmentation of large a-C:H/a-C grains is poorly understood, but the last panel of Figure \ref{ill_fam_Md} demonstrates a large value of $k_{\rm frag}$ results in a reduction in the dust content of low-metallicity galaxies. This could provide an alternative destruction mechanism for explaining the low dust content of low-metallicity galaxies.

\section{Statistical constraints from MCMC}
\label{MCMCresults}

\subsection{Statistical constraints on metal parameters}
\label{MCMCresults_Z}

\begin{figure*}
  \center
\includegraphics[width=0.9\textwidth]{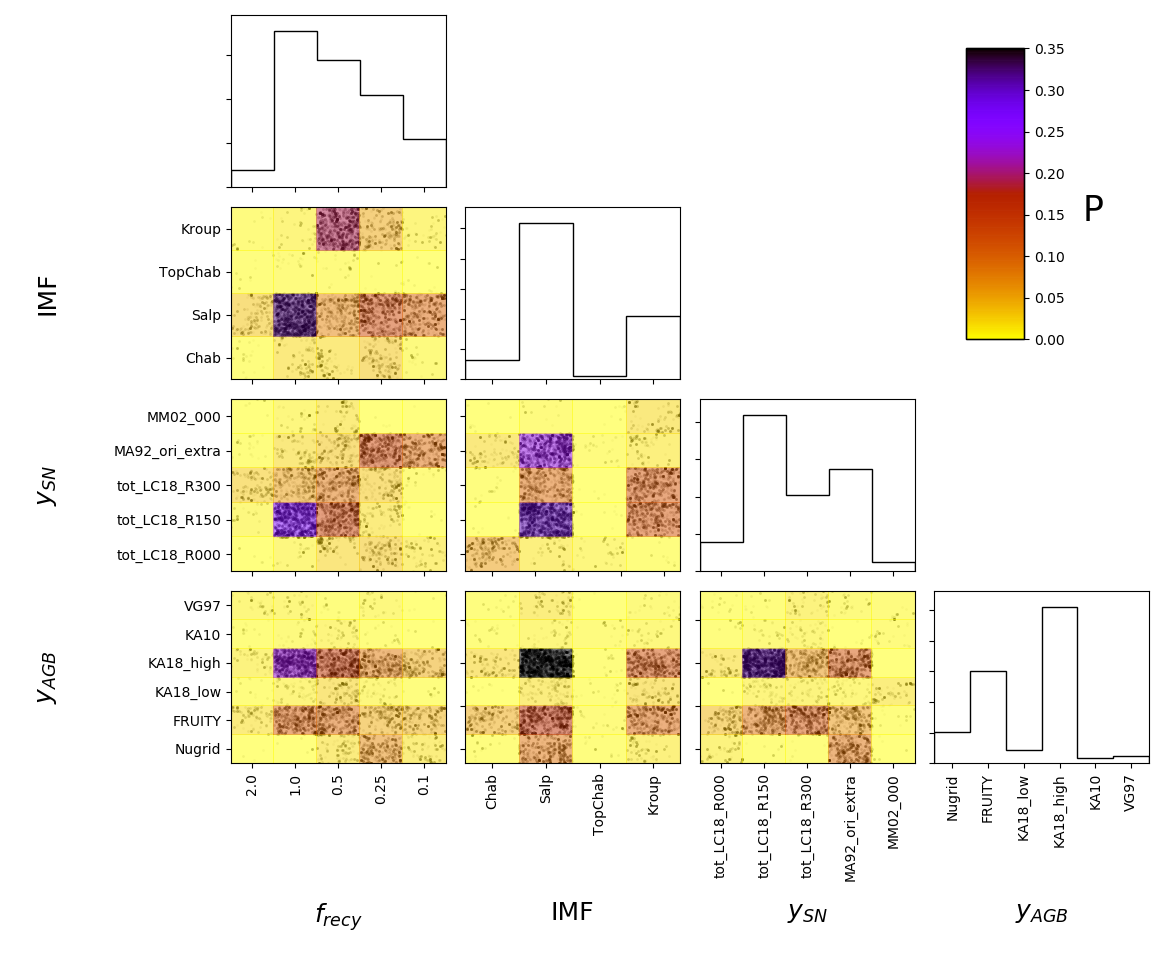}
  \caption{Corner plot showing the probability distributions (see colourbar) on the grid used for studying the metallicity-related parameters. The probability distributions reveal multiple likely combinations (i.e. degeneracies).}
  \label{cornerZ}
\end{figure*}

By comparing our grid of chemical evolution models to all available observations simultaneously, we obtain probability functions for each of the four input parameters (IMF, $mp_{Z,\textrm{SN}}$, $mp_{Z,\textrm{AGB}}$ and $f_{\rm recy}$). Additionally, it is also possible to make 2D probability distributions between any two of the input parameters. The corner plot showing these probability distributions is shown in Figure \ref{cornerZ}.
From the histograms, it is immediately clear that some input-values in our grid do not result in realistic models. Decreasing the recycling time scaling factor $f_{\rm recy}$ by a factor of 10, or increasing it by a factor of 2 results in statistically poor fits. Similarly a Chabrier IMF or Top-heavy Chabrier IMF have low probability, as well as SN yields of \citet{Meynet2002} or the non-rotating models of LC18. The AGB-yields of \citet{vandenHoek1997} and \citet{Karakas2010} also have low probability. 

We now study the 2D probability distributions of the MCMC trace in further detail. We find that the most likely combination of parameters is the model with $f_{\rm recy}=1$, Salpeter IMF, LC18 SN yields with $v_{\rm rot}=150\ \rm{km/s}$ and AGB yields from \citet{Karakas2018} with high mass loss rates. Another particularly good combination of parameter values is found for the model with $f_{\rm recy}=0.25/0.1$, Salpeter IMF, \citet{Maeder1992} SN yields and AGB yields from \citet{Karakas2018} with high mass loss rates. There is a third option that also has non-negligible probability, with a Kroupa IMF, $f_{\rm recy}=0.5$, LC18 SN yields with $v_{\rm rot}=150\ \rm{km/s}$ and AGB yields from \citet{Karakas2018} with high mass loss rates. In addition to these three most likely options, there are additional variations that give non-neglibiible contributions. 

There are thus three groups of models that each have entirely different combinations of IMF, $f_{\rm recy}$ and SN yield tables yet produce probable matches to the observations in spite of their different numerical values. This demonstrates a degeneracy between the IMF, SN yields and $f_{\rm recy}$. Using the 2D histograms is paramount for identifying the most likely parameter-combinations and revealing degeneracies.

\subsection{Statistical constraints on dust parameters}
\label{MCMCresults_Md}
\begin{figure*}
  \center
\includegraphics[width=0.9\textwidth]{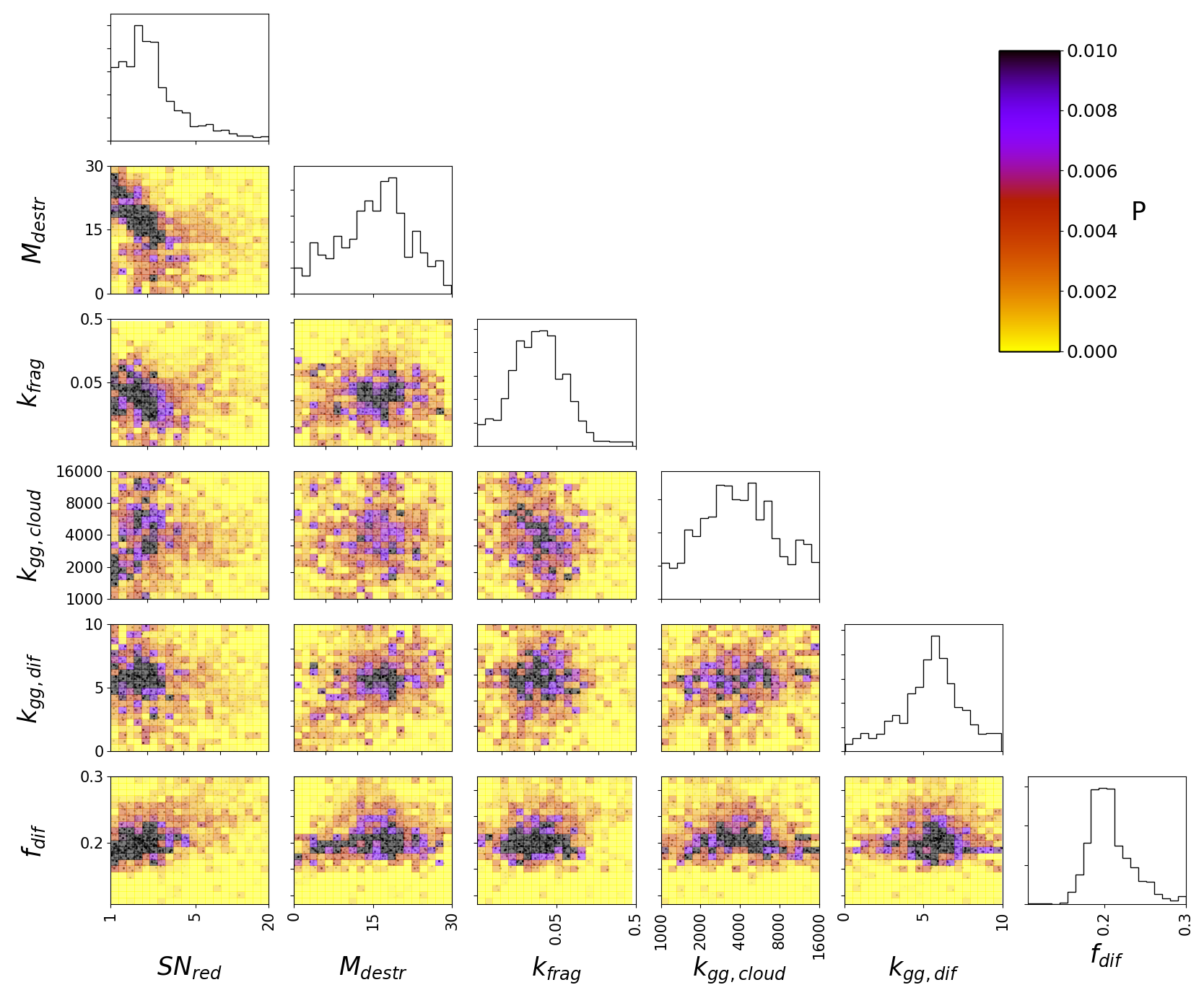}
  \caption{Corner plot showing the probability distributions (probability indicated by the colourbar) for the continuous MCMC run studying the dust-related parameters.}
  \label{cornerMd_cont}
\end{figure*}

\begin{table}
\caption{Best values and uncertainties on the dust parameters from the percentiles of their distribution in the continuous MCMC sample.}
\centering
\footnotesize
\begin{tabular}{l|lll} \hline\hline
Parameter & Median & $\rm 84^{th}-50^{th}$ & $\rm 16^{th}-50^{th}$ \\ 
 &  &  Percentile & Percentile \\ \hline
$\rm SN_{\rm red}$ & 2.01 & 2.21 & -0.72 \\
$M_{\rm destr}$ & 15.74 $M_\odot$ & 6.54 $M_\odot$ & -8.06 $M_\odot$ \\
$k_{\rm frag}$ &  0.030 & 0.039 & -0.016 \\
$k_{\rm gg,cloud}$ &  3820 & 4393 & -1886 \\
$k_{\rm gg,dif}$ & 5.59 $\rm Gyr^{-1}$ & 1.72 $\rm Gyr^{-1}$ & -2.27 $\rm Gyr^{-1}$\\
$f_{\rm dif}$ & 0.204 & 0.035 & -0.020 \\ 
 \hline \hline
\end{tabular}
\label{tab:best_fits}
\end{table}

Figure \ref{cornerMd_cont} shows the corner plot with the 1D and 2D probability distributions for the continuous MCMC run constraining the dust parameters.
Allowing the dust parameters to vary continuously results in relatively well-behaved Gaussian probabilty distributions. The MCMC trace is able to find the most likely parameter values and iterate around the most likely values around it. The results are listed in Table \ref{tab:best_fits}.  The uncertainties on most of the parameters are on the order of 50\,per\,cent. The only parameter that is strongly constrained is the fraction of metals available for grain growth in the diffuse ISM, $f_{\rm dif}$. The resulting value is consistent with the dust-to-metal ratio of $M_d/M_Z=0.214$ for DustPedia as measured by \citet{DeVis2019}.

Next, we look at the degeneracies between the different parameters. There are multiple positive and negative correlations between the parameters. The strongest positive correlations observed are (i) the correlation between $\rm SN_{red}$ and $f_{\rm dif}$, and (ii) between $k_{\rm gg,dif}$ and $M_{\rm destr}$. These positive correlation can be explained as the effects of these parameters cancel each-other out. One of the terms increases the dust content (increasing $f_{\rm dif}$ or $k_{\rm gg,dif}$) and the other reduces the dust content (increasing $\rm SN_{red}$ or $M_{\rm destr}$). Figure \ref{ill_fam_Md} demonstrates that $k_{\rm gg,dif}$ and $M_{\rm destr}$ have very similar, but opposite, effects.

The strongest negative correlation is between $\rm SN_{red}$ and $M_{\rm destr}$, both of which decrease the dust content.  Finally, there is also an anti-correlation between $\rm SN_{red}$ and $k_{\rm frag}$ (especially if the values with $\rm SN_{ red}>5$ and $k_{\rm frag}>0.1$ are ignored). The latter correlation arises from the constraints imposed by the low metallicity galaxies in our sample. In order to obtain a good fit to these unevolved galaxies with our chemical evolution models, $\rm SN_{red}$ or $k_{\rm frag}$ have to be increased (Figure \ref{ill_fam_Md}). Both reduced SN dust or photo-fragmentation can explain the low dust masses of low metallicity galaxies.

\subsection{Dust budget of best models}
\begin{figure*}
  \center
\includegraphics[width=0.45\textwidth]{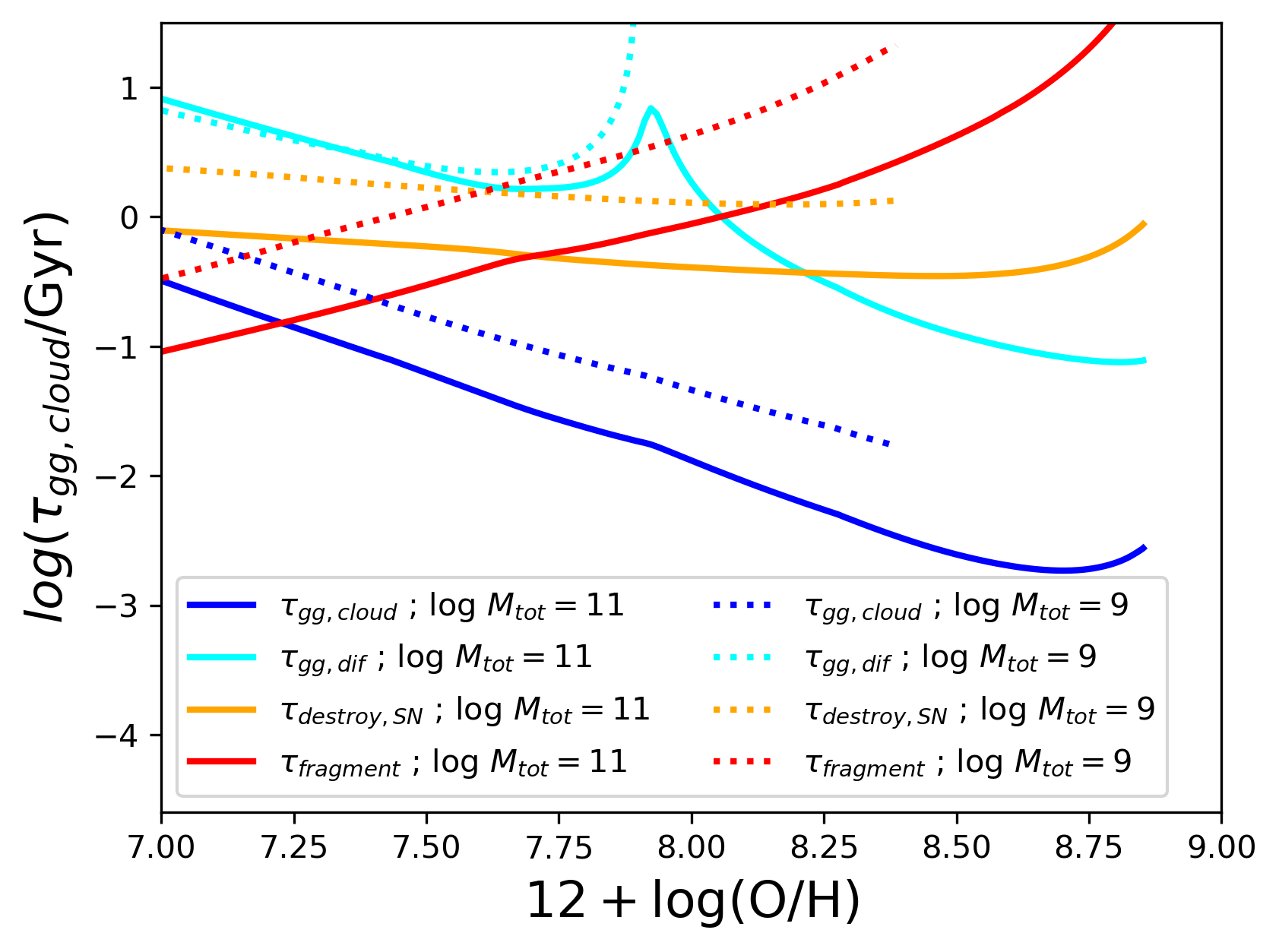}
\hspace{0.03\textwidth}
\includegraphics[width=0.45\textwidth]{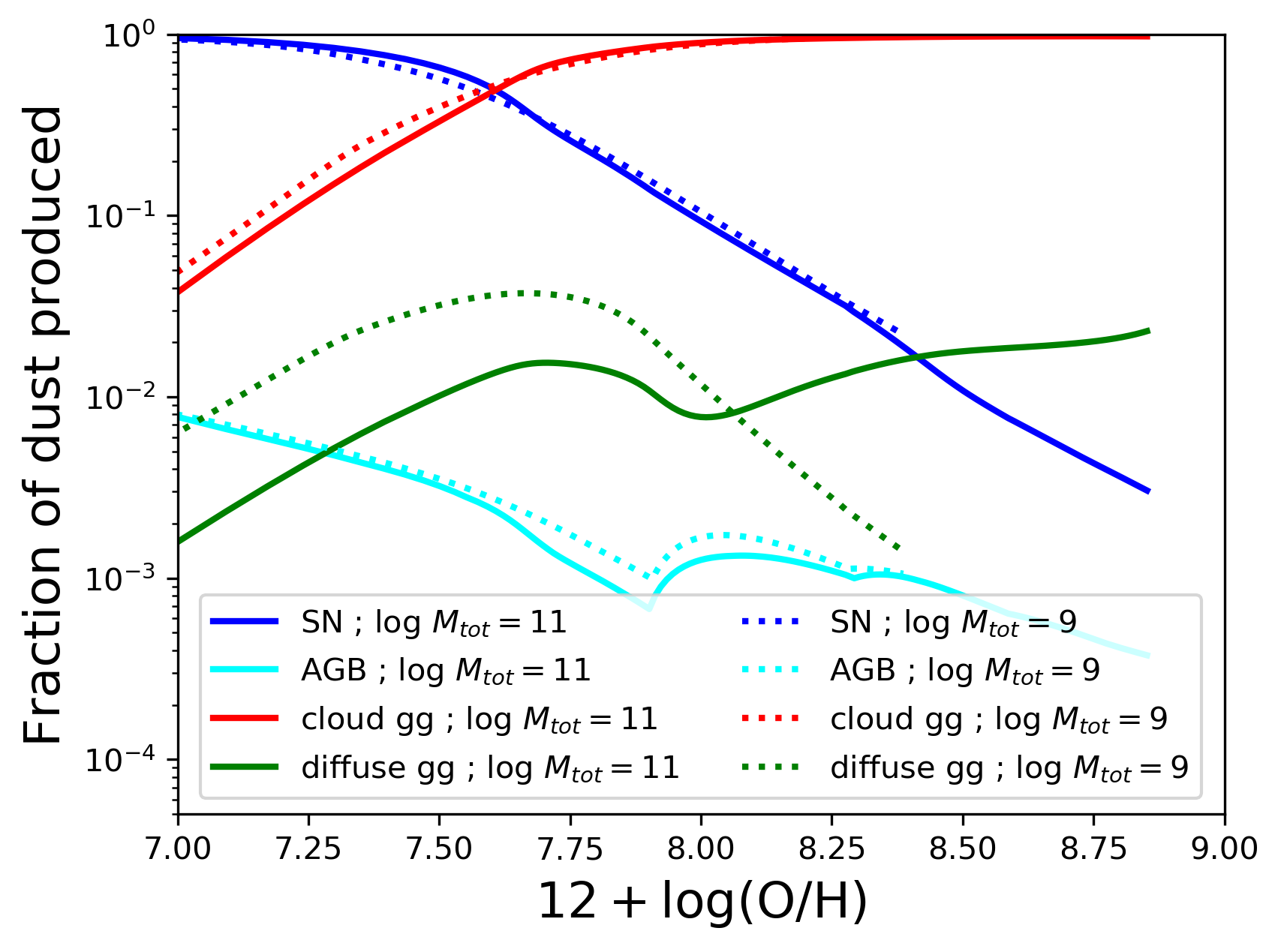}
  \caption{\textit{Left:} Evolution in the timescales associated with the different dust processing mechanisms with increasing metallicity. These include grain growth in the diffuse ISM (cyan) and dense ISM (blue), dust destruction in SNe (orange) and photo-fragmentation (red).  Solid and dotted lines indicate different $M_{\rm tot}$ values. \textit{Right:}  Similar to the \textit{left} image, but now following the fraction of dust mass contributed by different dust sources including SNe (blue), AGB stars (cyan), grain growth in the dense (red) and diffuse ISM (green). Galaxies transition from being SN-dust dominated to cloud grain growth dust dominated around a metallicity of $12+\log({\rm O/H})=7.75$.  }
  \label{bestmodel}
\end{figure*}

To improve on our interpretation, we look in more detail at the dust budget of our best `family' of models. Here we have run models with different initial masses and SFE and taken the best values from Table \ref{tab:best_fits} as the relevant dust parameters. Figure \ref{bestmodel} (left) shows how the timescales associated with the different dust processing mechanisms vary with metallicity for two different galaxy masses. We see that for our best models with $M_{tot}=10^9M_\odot$, the cloud dust grain growth timescale varies on a linear power-law from about 1 billion years for low metallicities till about 10 million years for high metallicities. The diffuse grain growth is over an order of magnitude slower and initially also follows a power law with metallicity. However towards high metallicities, the diffuse grain growth first levels off as the maximum dust-to-metal ratio in the diffuse ISM is reached (which reduces the diffuse grain growth timescale as apparent in Equation \ref{ggcloudeqn}) and then increases again as a balance is formed between dust production and dust destruction.

The dust destruction by SN shocks is initially inefficient due to the low dust-to-gas ratios in the low metallicity ISM. As the dust-to-gas ratio increases with increasing metallicity, the dust destruction timescales for SN shocks become faster, but it never catches up with the increasing rate of the dust grain growth. The photo-fragmentation of large grains is the dominating process (i.e. short timescales) at low metallicities and then becomes weaker (longer timescales) as the galaxy evolves. This can be explained in the following way: in unevolved galaxies, there are many young stars and thus there is a very harsh radiation field. In addition, due to the low dust-to-gas ratio the dust does not self-shield and the photo-fragmentation of dust grains is thus initially very efficient. As galaxies evolve the radiation field becomes less harsh and the large dust grains can survive longer. 

Next we look at how much the different dust production mechanisms contribute to the total produced dust mass (i.e. ignoring all dust destruction) in Figure \ref{bestmodel} (right). We see that at low metallicities, SN contribute nearly all of the dust mass, and AGB stars and grain growth only produce a marginal fraction. However as the metallicity increases, the efficiency of both the cloud and diffuse grain growth increases. At a metallicity of about $\rm 12+\log(O/H)=7.75$, cloud grain growth has produced half of the dust present in the galaxy, and from then on cloud grain growth is the dominant mechanism. This is a significantly lower transition value than for other models in the literature \citep{Kuo2012,Vilchez2019,DeVis2019}, we note that these studies did not include photo-fragmentation.

In order to explain the low dust-to-gas and dust-to-metal ratios observed in multiple observational studies (\citealp{Remy-Ruyer2014,DeVis2017a,DeVis2019}; Cigan et al, {\it in preparation}), previous modelling attempts \citep[e.g][]{Zhukovska2014,Feldmann2015,DeVis2017b} have often invoked very low SN dust yields. In this work we have found that including a dust destruction term for photo-fragmentation of grains can also explain the lower dust-to-gas and dust-to-metal ratios of low metallicity galaxies. In this case, the SN dust contribution only needs to be lowered by a factor of 2 to about on average $\sim 0.5~M_\odot$ of dust per SN. This is more consistent with observations of SN remnants \citep{Dunne2003,Morgan2003B,Rho2008,Barlow2010,Matsuura2011,Matsuura2015,Temim2017,Rho2018,Chawner2019} than the more extreme values that are necessary without photo-fragmentation of large grains.

\section{Caveats}
\label{sec:caveat}
Finally we want to point out a number of important caveats to this work. The first and most important being that we do not claim our model can be used to state that a given parameter (e.g. Salpeter vs Chabrier IMF) is definitively more realistic than another. Within our statistical framework, one parameter might give a better fit than another and this does count for something, but this is only valid within the assumptions that were made (from the assumption of the outflow prescription to the dust mass absorption coefficient). There are a number of additional factors that need to be considered. 

The first consideration is that we have only modeled galaxies evolving in isolation, and have not taken into account their merging histories. In addition, the prescription we used for inflows and outflows might not be ideal for our type of galaxies. The \citet{Nelson2019} prescription was chosen as the most realistic available, but may not encompass the physical properties of the whole range of galaxies observed in our sample. Additionally, when calculating the redshifts for the outflows, it is assumed that each galaxy was formed shortly after the Big Bang. When comparing our models to the observations, we allow the age of the galaxy to vary (i.e. we compare to the model tracks rather than just the model end point). This means we are effectively allowing the formation time to be much later than the Big Bang. For models where this is the case, the model will have used outflows based on a redshift that is
slightly too large. This introduces an error into our models, but we
consider this error to be negligible compared to the overall uncertainty
associated with the outflow prescription.

In Section \ref{MCMCresults_Z}, we found there is a degeneracy between the IMF, $f_{\rm recy}$ and SN yield tables. Rather than varying $f_{\rm recy}$, it is also possible to change the outflow prescription. This strongly affects the degeneracy with IMF and SN yield tables. In the preparation of this study, multiple outflow prescriptions were tried, with significantly different results for what parameters gave the best fit (e.g. the Chabrier IMF was found as best IMF in one iteration). However, even though the results were different,  a very similar degeneracy between IMF, SN yield tables, and outflows was always recovered. 

Another key caveat is that our `one-zone' approach without resolution has inherent limitations. We do not resolve different regions within the galaxy (other than the separation of the diffuse and cloud ISM). A real galaxy does, of course, not evolve so uniformly and different regions evolve at different timescales. This undoubtedly affects the chemical evolution of galaxies. Detailed comparisons between resolved hydrodynamical simulations (which can unfortunately not sample the dust processing parameter space in detail) and `one-zone' models could shed light on where the shortcomings of the latter lie, and how they can be overcome.

In addition, in this work we have only considered large dust grains without separating between silicate and carbonaceous dust. Silicate and carbonaceous dust do not have the same origins nor the same evolutionary pathways. Furthermore, the carbonaceous dust component should ideally include information about the grain size distribution or be divided into two populations. There would then be one population with silicate dust (not sensitive to photo-processing), one population with large a-C(:H) grains (prone to photo-fragmentation), and one with nano a-C grains (subject to photo-dissociative processing into their hydrocarbon or atomic constituents).
Note that the two latter destructive routes are energetic EUV photon-driven but the 
microscopic processes are different and depend on the ISRF (photo-fragmentation, important at the grain sub-structure level, versus photo-dissociation at the C-H and C-C bond molecular level; \citeb{Schirmer2020}).
Including these different populations could lead to a more detailed understanding of dust evolution (especially when combined with MIR-based observational constraints).

Another consideration is that the priors we have used might not be the most appropriate for this kind of work. Constraints from theoretical modelling could improve our prior knowledge of the possible parameter values, which could in turn change our MCMC results. By using a uniform prior (either in logarithmic or linear space depending on the kind of parameter), we have aimed to use the most uninformative prior available, yet improvements could be made in future work by constructing realistic priors consistent with theoretical predictions.

Finally, a further improvement could also be made to our statistical framework. Currently we have added measurement and model uncertainties, where the model uncertainty is to account for systematic errors and the limited model sampling. However in our approach these uncertainties are treated as random uncertainties, and thus scale as $\sigma \propto \sqrt{N}$ (where $N$ is the number of observations). In reality, the systematic uncertainty component would result in the same error for each galaxy (e.g. if the dust mass absorption coefficient is biased, this bias would be the same for each galaxy) and would thus not scale with the number of observations. To account for this in our framework, we would need to use a full error co-variance matrix in the likelihood calculation in Equation \ref{Eq:MCMC5}. Unfortunately, this would make this calculation too time consuming for the number of galaxies, models and computational power we are using. 

Instead, we have tried to address some of the effects of systematic biases by manually perturbing all the observations. We added a bias of 0.1 dex to each of the observed variables one-by-one, and repeated our analysis. For brevity, we do not show all these results, but only summarise that these perturbations had only a limited effect on the probability distributions and did not affect any of the conclusions in this work. We also experimented with changing the metallicity calibration, where we use the IZI metallicities from \citet{DeVis2019} instead of PG16S. The results using the IZI calibration are shown in Appendix \ref{IZI}.
These kind of perturbations can give us some idea of the systematic effects. However, doing a more realistic calculation including a full covariance matrix, would make the uncertainties on the chemical and dust evolution parameters significantly more realistic, and would be the logical next step. 

\section{Conclusions}
\label{sec:concl}
In this work we have used a rigorous statistical framework to provide statistical constraints on the chemical and dust evolution parameters for nearby galaxies. The models are compared to 340 nearby late-type  galaxy observations from the DustPedia, HIGH and HAPLESS samples. The key effects of varying each of the parameters is illustrated using plots of how the metallicity increases with decreasing gas fraction, and how the dust-to-gas ratio increases with increasing metallicity. A Bayesian MCMC framework was used to provide statistical constraints, where the relative probabilities were calculated from the $\chi^2$, and we marginalised over the different time steps, 10 different galaxy masses and 6 star formation histories. Metallicity-related parameters were studied first, and the best model was chosen for subsequently studying the dust parameters. From studying the statistical constraints we conclude:
\begin{itemize}
    \item Our main conclusion from exploring the full parameter space is that there are multiple viable models that compare well to the observations. In other words, there are significant degeneracies between various chemical and dust evolution parameters. It is thus necessary to sample a large parameter-space when trying to put statistical constraints on the model parameters.

    \item There are multiple combinations of metallicity parameters that give a realistic fit to the build up of metals as galaxies evolve. In particular, there is a degeneracy between varying the IMF, $f_{\rm recy}$ and SN yield tables. The best fitting combination is the one with a Salpeter IMF, $f_{\rm recy}=1$ and \citet{Limongi2018} SN yields with $v_{rot}=150\ \rm{km/s}$. We note this fit is particularly dependent on the outflow prescription.
    
    \item The \citet{Karakas2018} AGB yields with high mass loss rates give consistently the best fitting results within our statistical framework. Older AGB yield tables such as the ones from \citet{vandenHoek1997} and \citet{Karakas2010} provide a poor fit to the build up of the nitrogen-to-oxygen ratio as galaxies evolve.
    
    \item From varying the dust parameters continuously, it is possible to find the best fitting values and uncertainties. The best fitting model has $SN_{\rm red} = 2.01$, $M_{\rm destr}= 15.74 M_\odot$, $k_{\rm frag}=0.030$, $k_{\rm gg,cloud} = 3820$, $k_{\rm gg,dif} = 5.59 \rm Gyr^{-1}$ and $f_{\rm dif} = 0.204$. The uncertainties on these parameters are typically quite large (on the order of 50\,per\,cent or more), except for $f_{\rm dif}$.
    
    \item There are degeneracies between a number of dust parameters. In particular, there is a positive correlation between $\rm SN_{red}$ and $f_{\rm dif}$ and between $k_{\rm gg,dif}$ and $M_{\rm destr}$. There is also a negative correlation between $\rm SN_{red}$ and $M_{\rm destr}$, and between $\rm SN_{red}$ and $k_{\rm frag}$. In other words, both SN dust reduction or photo-fragmentation can explain the low dust content of unevolved galaxies. The best fit is for a combination of the two.
    
    \item For the best fitting models, the grain growth timescales get shorter following a linear power-law with increasing metallicity. The photo-fragmentation timescales get longer with increasing metallicity. Nearly all the dust of low-metallicity galaxies was made by SN. The dust mass of high metallicity galaxies is dominated by grain growth dust. Galaxies with PG16 metallicities around $12+\log({\rm O/H})=7.75$ have similar amounts of SN and grain growth dust. This is a lower transition metallicity than most models in the literature.
    
\end{itemize}

\section*{Data availability}
The DustPedia data used in this paper are available from the DustPedia Archive (http://dustpedia.astro.noa.gr/) and the HIGH and HAPLESS data are tabulated in \citet{DeVis2017a} and \citet{Clark2015} respectively. The chemical evolution code used for producing all the chemical evolution models in this work is publicly available on Github (https://github.com/zemogle/chemevol).

\section*{Acknowledgments}
The authors gratefully acknowledge Mercedes Molla for providing their stellar metal yield tables and Dylan Nelson for providing outflow rates and velocities of the different outflow components as well as interesting discussion. PDV, SJM, HLG and LD acknowledge support from the European Research Council (ERC) in the form of Consolidator Grant {\sc CosmicDust}.  

\bibliographystyle{mnras}
\bibliography{Library}

\appendix

\section{Results using IZI metallicities}
\label{IZI}
\begin{figure*}
  \center
\includegraphics[width=0.7\textwidth]{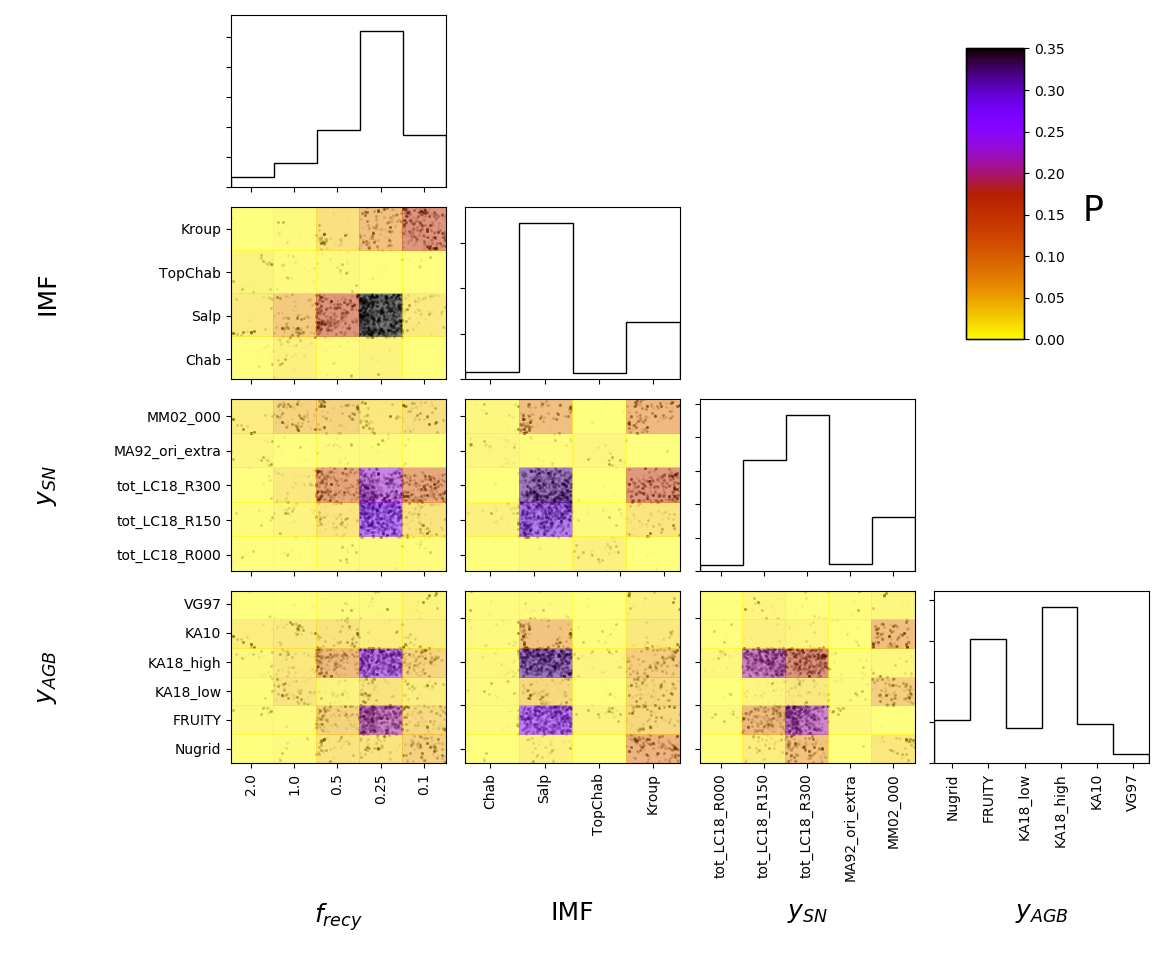}
  \caption{Corner plot showing the probability distributions on the grid used for studying the metallicity-related parameters when using the IZI metallicity calibration instead of PG16 (see Figure \ref{cornerZ}). The results are different than for PG16, but all main conclusions hold.}
  \label{cornerZ_IZI}
\end{figure*}

In this appendix, we show results for repeating our analysis using IZI metallicities from \citet{DeVis2019} instead of PG16S. The corner plot for the metal grid is given in Figure \ref{cornerZ_IZI} shows the statistical constraints. We see that the results are different than for PG16S, with e.g. the best scaling factor for the outflow recycling timescale now being $f_{\rm recy}=0.25$. However we can still see that there are multiple combinations of IMF, SN-yield and recycling time factor that give a reasonable fit. Our main conclusion that there is a degeneracy between these parameters thus holds.

\begin{figure*}
  \center
\includegraphics[width=0.7\textwidth]{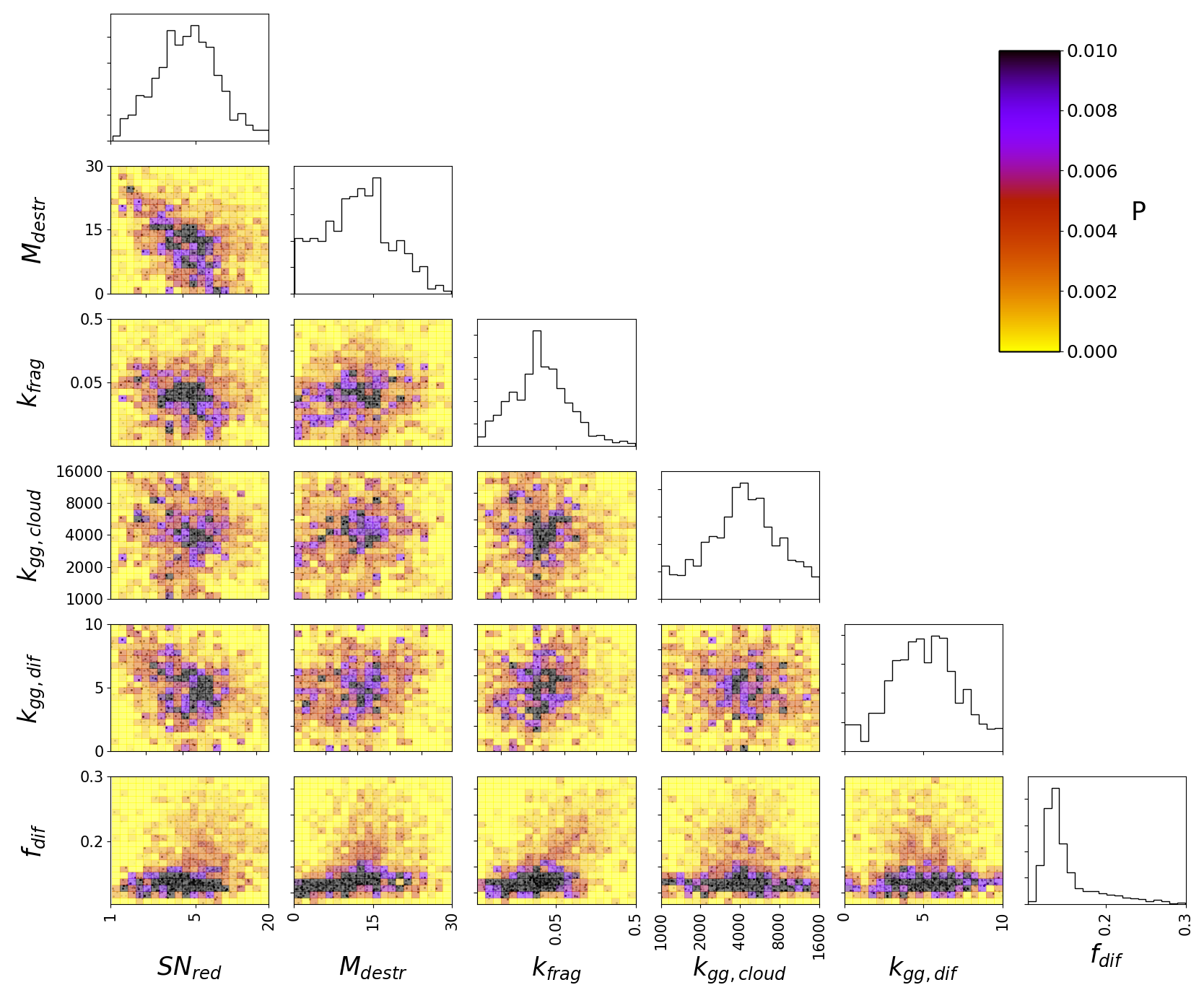}
  \caption{Corner plot showing the probability distributions for the dust parameters when using the IZI metallicity calibration instead of PG16 (see Figure \ref{cornerMd_cont}). The results are similar to those for PG16, with the main difference being lower $f_{\rm dif}$.}
  \label{cornerMd_IZI}
\end{figure*}

We then used the best results from the IZI metal grid to constrain the dust parameters, following our approach detailed in Section \ref{MCMC}. The resulting corner plot for the dust parameters is given in Figure \ref{cornerMd_IZI}. Overall the results are quite similar to those using PG16. The biggest difference is that $f_{\rm dif}$ is significantly lower for the IZI results ($f_{\rm dif} = 0.139^{+0.042}_{-0.014}$). This difference is entirely expected as changing the metallicity calibration will directly affect the dust-to-metal ratios. We again find that $f_{\rm dif}$ is entirely consistent with the average dust-to-metal ratio of 0.141 for the DustPedia sample using the IZI calibration \citep{DeVis2019}. The other parameter that is affected is $\rm SN_{red}$, which has increased to $4.43^{+3.08}_{-1.96}$ (i.e. to about $1\sigma$ from the PG16 value). This is not very surprising though since we already knew from Section \ref{MCMCresults_Md} that there is a correlation between $f_{\rm dif}$ and $\rm SN_{red}$. The other results are not much affected, and we find similar degeneracies between the various parameters (e.g. between $\rm SN_{red}$ and $f_{\rm dif}$ or between $\rm SN_{red}$ and $k_{\rm frag}$). Our conclusions are thus not sensitive to the choice of metallicity calibration.

\label{lastpage}
\end{document}